\RequirePackage{fix-cm}
\documentclass[smallextended]{svjour3}       
\smartqed  
\usepackage{graphicx}
\usepackage{bm}
\usepackage{float}
\usepackage{framed}
\usepackage{amsfonts, amsmath, amssymb, amsbsy}
\usepackage{appendix}
\usepackage{hyperref}
\usepackage{cite}
\usepackage[caption=false, font=normalsize, labelfont=sf, textfont=sf]{subfig}
\allowdisplaybreaks[4]
\usepackage{xcolor}
\definecolor{lzcol}{rgb}{1, 0, 0}

\definecolor{yscol}{HTML}{6622AA}

\newtheorem{assumption}{Assumption}
\newenvironment{customproof}[1]
{\noindent\textit{Proof of Remark~\ref{#1}.}}

\begin{document}

\title{Amplitude Expansion Phase Field Crystal (APFC) Modeling based Efficient Dislocation Simulations using Fourier Pseudospectral Method}

\titlerunning{APFC Dislocation Simulations using Fourier PSM}        

\author{Xinyi Wei \and Yangshuai Wang \and Kai Jiang \and Lei Zhang }


\institute{Xinyi Wei\at
          School of Mathematical Sciences, Institute of Natural Sciences and MOE-LSC, Shanghai Jiao Tong University, Shanghai 200240, China\\
          \email{xinyiwei@sjtu.edu.cn}
       \and
       Yangshuai Wang\at
          Department of Mathematics, Faculty of Science, National University of Singapore, 10 Lower Kent Ridge Road, Singapore\\
          \email{yswang@nus.edu.sg}
       \and
       Kai Jiang\at
          Hunan Key Laboratory for Computation and Simulation in Science and Engineering, Key Laboratory of Intelligent Computing and Information Processing of Ministry of Education, School of Mathematics and Computational Science, Xiangtan University, Xiangtan, Hunan, 411105, China\\
          \email{kaijiang@xtu.edu.cn}
       \and
       Lei Zhang\at
          School of Mathematical Sciences, Institute of Natural Sciences and MOE-LSC, Shanghai Jiao Tong University, Shanghai 200240, China\\
          \email{lzhang2012@sjtu.edu.cn}
}

\date{Received: date / Accepted: date}

\maketitle

\begin{abstract}
Crystalline defects critically influence material properties, necessitating accurate simulation methods. Existing approaches, from atomic-scale configurations to continuum elasticity, face inherent limitations in modeling dislocation-induced lattice deformation. The amplitude expansion of the phase field crystal (APFC) model bridges this gap with a mesoscopic description. This paper introduces a computationally efficient Fourier pseudospectral method for solving the APFC equations. The method exploits system periodicity and solution analyticity—the latter's rigorous proof remaining an open question, as discussed herein—to enable precise implementation of periodic boundary conditions. Numerical experiments on 2D triangular and 3D body-centered cubic lattices demonstrate that the method accurately reproduces the strain fields of edge dislocations, matching continuum theory predictions. These results confirm the APFC model's potential for capturing complex defect structures at the mesoscale, paving the way for simulating more intricate defect dynamics.

\keywords{Crystal defects \and Amplitude expansion \and Phase field crystal model \and Fourier pseudospectral method \and Elastic strain field}
\subclass{65M70 \and 65Z05 \and 74E15}
\end{abstract}

\section{Introduction}
\label{sec:introduction}

Crystalline defects, including point defects, dislocations, and cracks, significantly influence the properties of crystalline solids~\cite{Lagerlof2018Crystal,James2017Deformation,wang2024theoretical, braun2022higher,chen2022qm}. Accurate and efficient computational methods are crucial for studying these defects, particularly dislocations, which are key to understanding material behavior. While microscopic simulations are often computationally intensive and macroscopic models may lack precision, mesoscopic approaches strike an effective balance by offering coarse-grained degrees of freedom that enhance computational efficiency while maintaining sufficient accuracy, making them especially suitable for dislocation simulations. Prominent mesoscopic models include discrete dislocation dynamics (DDD)~\cite{lepinoux1987dynamic,ghoniem1988computer}, coarse-grained molecular dynamics (CGMD)~\cite{rudd1998coarse,joshi2021review}, kinetic Monte Carlo (KMC)~\cite{lin1999kinetic,cai2001kinetic}, and phase field crystal (PFC)~\cite{Elder2007Phase-field,Stefanovic2006Phase-Field}.

The PFC model excels in describing changes in the free energy functional with respect to local atomistic density, providing valuable insights into elasticity and dislocation dynamics. However, its application is limited by the need for fine spatial discretization. To address this, Goldenfeld et al.~\cite{PhysRevE.72.020601} introduced the complex amplitude expansion of the PFC (APFC) model. This extension, based on the plane wave expansion of the density, allows for the exploration of longer time scales and larger length scales, significantly broadening the model’s applicability in advanced material modeling~\cite{ofori2013complex,ElderAmplitude2010,jreidini2021orientation}. The APFC model is particularly useful for studying elastic effects~\cite{PhysRevE.89.032411,ElderAmplitude2010}, dislocation dynamics~\cite{PhysRevB.97.054113,PhysRevLett.121.255501}, and the control of material properties~\cite{2017Controlling, Choudhary01092012}. It can be effectively applied across various atomic descriptions, including those from theoretical modeling~\cite{de2023amplitude}, atomic simulations~\cite{wang2023pattern}, and experimental imaging~\cite{ElderAmplitude2010}.

The primary challenge in APFC models lies in the development and implementation of efficient numerical algorithms, particularly for dislocation simulations, which are crucial in computational materials science. The model is nonlinear with higher-order terms, and the corresponding dynamic equations involve multiple amplitude functions associated with reciprocal lattice vectors. Various numerical methodologies have been proposed to solve the APFC model. The finite difference method was initially employed~\cite{PhysRevE.72.020601,Yeon07012010}, but it suffered from instability issues, requiring constraints on the time step due to grid spacing limitations~\cite{provatas2011phase}. To address this difficulty, semi-implicit methods and spatial adaptivity have been explored. For example, finite element method~\cite{2017Controlling, 2015Software,PhysRevE.76.056706,PhysRevE.98.033303} and Fourier spectral method~\cite{PhysRevB.81.214201,hirvonen2016multiscale} have been utilized for APFC models. However, finite element methods may not optimally exploit the periodic lattice structures in crystal models. While the Fourier spectral method offers a global representation through Fourier series, it can result in increased free energy if the time step is too large for certain models~\cite{hirvonen2016multiscale}. Additionally, efficiency remains a concern across these approaches. Therefore, further research is required to comprehensively address these challenges.

The Fourier pseudospectral method~\cite{2011Spectral,gottlieb1977numerical} solves PDEs at collocation points, simplifying the treatment of nonlinear terms. Its computational efficiency, scaling as $\mathcal{O}(N\log N)$ via the fast Fourier transform, is ideal for leveraging the periodicity of crystalline structures. We apply this method spatially, combined with a semi-implicit time integration scheme, to solve the APFC model for edge dislocations in 2D triangular and 3D BCC lattices. With a suitable initial guess, we compute amplitude functions, from which the derived strain fields align with theoretical predictions~\cite{anderson2017theory}. We further validate the results by analyzing the long-range decay of the strain, showing consistency with continuum elasticity. We note that the spectral convergence observed numerically in Section~\ref{sec:numerical experiments} requires a rigorous justification. Such a proof would depend on the analyticity of the exact solution (Assumption \ref{assumption}) and likely the energy dissipation of both the continuous and discrete dynamics. This analysis is beyond the scope of this paper and will be addressed in future work (see Remark~\ref{remark:analyticity} and Remark~\ref{remark:error-estimate}).

This paper primarily focuses on applying the Fourier pseudospectral method within the APFC model to simulate dislocations, highlighting its potential and advantages in solving periodic crystalline defect systems. However, the proposed analysis and methodology can also be extended to grain boundaries and other complex scenarios. We plan to explore the generalization of these approaches in future research and discuss potential developments in Section~\ref{sec:conclusion}.

\subsection*{Outline} 
This paper is organized as follows: Section~\ref{sec:APFC Model} introduces the APFC model and the lattice configurations required for numerical simulations. Section~\ref{sec:numerical methods} details the Fourier pseudospectral method, including its formulation and the effective implementation for solving the APFC model. Section~\ref{sec:numerical experiments} presents numerical results for edge dislocations in both two-dimensional triangular and three-dimensional body-centered cubic lattices, with analysis and interpretation within the context of the APFC model. Section~\ref{sec:conclusion} summarizes the key findings and outlines future research directions. The appendices provide supplementary information for interested readers.

\section{APFC Model}
\label{sec:APFC Model}

In this section, we provide a derivation of the APFC model from the PFC free energy functional, as well as the corresponding evolution equations for the crystalline lattices considered in this work. The PFC free energy functional ~\cite{Elder2007Phase-field,Stefanovic2006Phase-Field} is defined with respect to the dimensionless atomic probability density $n$, which serves as a continuum representation of discrete atomic lattices. The functional is given by
\begin{equation}
	F_n = \frac{1}{\lvert\Omega\rvert}\int_{\Omega} \Big[\frac{B_0}{2} n^2 + \frac{B_x}{2}n(1+ \nabla ^2 )^2 n-\frac{\gamma}{3}n^3 + \frac{v}{4}n^4\Big] \,{\rm d}\bm{r},
	\label{eqn:energyfunctional}
\end{equation}
where $\Omega$ is the computational domain and $B_x \text{,}\; v\text{,}\;\gamma$, and $B_0$ are parameters that control the phase diagram and properties of the underlying system~\cite{Elder2007Phase-field}. The APFC model~\cite{2006Renormalization,Athreya2006Renormalization,PhysRevE.72.020601} can be derived by incorporating crystalline information and approximating the density $n$ as a sum of plane waves
\begin{equation}
	n(\bm{r},t) = n_0(\bm{r},t) + \sum_{j=1}^M \big[\eta_j(\bm{r},t)e^{\imath\bm{k}_j \cdot \bm{r}} + \eta_j ^* (\bm{r},t)e^{-\imath\bm{k}_j \cdot \bm{r}} \big].
	\label{eqn:apfcdensity}
\end{equation}
The set $\{\bm{k}_j\}_{j=1}^{M}$ consists of $M$ non-zero reciprocal lattice vectors that represent distinct lattice symmetries, defined as $\bm{k}_j = \sum_{i=1}^{M} n_i \bm{b}_i$, where $\bm{b}_i$ are the primitive reciprocal lattice vectors and $n_i \in \mathbb{Z}$ are their coefficients. They satisfy the relation $\bm{a}_i \cdot \bm{b}_j = 2\pi \delta_{ij}$, with $\bm{a}_i$ being the primitive vectors of the crystal lattice~\cite{zhang2008efficient}. The variables $\eta_j(\bm{r},t)$ are amplitude functions corresponding to each plane wave, which are crucial for characterizing the state and evolution of the crystal system. Without loss of generality, the average density $n_0(\bm{r},t)$ can be set to zero~\cite{ElderAmplitude2010}. We will illustrate two specific examples at the end of this section (cf. Example~\ref{ex}).

In the PFC model, the density $n$ satisfies the following dynamic equation, which represents a gradient flow of the energy functional
\begin{equation}
	\frac{\partial n}{\partial t} = \nabla^2\frac{\delta F_n}{\delta n},
	\label{eqn:pfcdynamics}
\end{equation}
where 
\begin{equation}
	\frac{\delta F_n}{\delta n} = B_0 n + B_x(1+\nabla^2)^2n - \gamma n^2 + vn^3.
\end{equation}

In the following, we derive the dynamic equation within the APFC framework. 

\paragraph{QDRG argument.}
Applying the $\nabla^2$ operator to $n$ and using \eqref{eqn:apfcdensity} yields
\begin{equation}
	\nabla^2n = \sum_{j}e^{\imath\bm{k}_j\cdot \bm{r}} (\nabla^2 + 2\imath\bm{k}_j\cdot\nabla - \lvert\bm{k}_j\rvert^2)\eta_j + \sum_{j}e^{-\imath\bm{k}_j\cdot \bm{r}} (\nabla^2 + 2\imath\bm{k}_j\cdot\nabla - \lvert\bm{k}_j\rvert^2)\eta^*_j.
	\label{long-wavelength limit}
\end{equation}
We then utilize the ``quick and dirty renormalization group" (QDRG) approach proposed by \cite{Athreya2006Renormalization} to capture the essential aspects of the calculations without relying on more rigorous methods. The main idea is to assume that the amplitudes $\eta_j(\bm{r},t)$ remain constant on atomic length scales. As shown in Remark~\ref{remark1}, this leads to
\begin{equation}
	\int_{C_1} \eta_j(\bm{r},t)e^{\imath\bm{k}_j\cdot {\bm{r}}} \,{\rm d}\bm{r} \thickapprox \eta_j(\bm{r},t)\int_{C_1} e^{\imath\bm{k}_j\cdot {\bm{r}}} \,{\rm d}\bm{r},
	\label{QDRG}
\end{equation}
where $C_1$ is the unit cell. Since $e^{\imath\bm{k}_j\cdot {\bm{r}}}$ is periodic in the crystal cell, the right-hand side of equation \eqref{QDRG} vanishes unless $\bm{k}_j = 0$.

\begin{remark}
	\label{remark1}
	To prove~\eqref{QDRG}, we consider the integral of its left-hand side over the cell $C_\varepsilon$ with $\varepsilon>0$. The following formula holds:
	\begin{equation}
		\Big\lvert \frac{1}{\varepsilon^d}\int_{C_{\varepsilon}}\eta(\bm{x},t)e^{\imath\bm{k}_j\cdot \frac{\bm{x}}{\varepsilon}} \,{\rm d}\bm{x} - \frac{1}{\varepsilon^d}\int_{C_{\varepsilon}}I_{\eta}e^{\imath\bm{k}_j\cdot \frac{\bm{x}}{\varepsilon}} \,{\rm d}\bm{x}\Big\rvert = O(\varepsilon),
		\label{eqn:remark1}
	\end{equation}
	where 
	\begin{equation}
		I_{\eta} = \frac{1}{\varepsilon^d}\int_{C_{\varepsilon}}\eta(\bm{x},t) \,{\rm d}\bm{x}
	\end{equation}
	represents the average value of the integral of $\eta(\bm{x},t)$ over $C_{\varepsilon}$.
	Here, $\varepsilon^d$ denotes the volume of the $C_{\varepsilon}$ cell, and $d$ is the dimension of the model. The complete proof is given in Appendix~\ref{apd:sub:proofs}.
\end{remark}

\paragraph{Dynamic equation of amplitude functions.} In the APFC framework, the dynamic equation for the density $ n $ given by \eqref{eqn:pfcdynamics} is transformed into a dynamic equation for the amplitude functions $ \eta_j $. This is achieved by multiplying the left-hand side of \eqref{eqn:pfcdynamics} by $ e^{-\imath \bm{k}_j \cdot \bm{r}} $, integrating over a unit cell $ C_0 $, and normalizing by its volume $ \lvert C_0\rvert $. By applying the QDRG approach, we can obtain 
\begin{align}
	\frac{1}{\lvert C_0\rvert}\int_{C_0} e^{-\imath\bm{k}_j\cdot {\bm{r}}}\cdot \frac{\partial n}{\partial t} \,{\rm d}\bm{r} &= \frac{1}{\lvert C_0\rvert}\int_{C_0} \frac{\partial\eta_j}{\partial t} + \sum\limits_{\substack{l=1\\ l\neq j}}^M\frac{\partial( \eta_le^{\imath(\bm{k}_l-\bm{k}_j) \cdot \bm{r}})}{\partial t} \,{\rm d}\bm{r}\\
	&\quad + \frac{1}{\lvert C_0\rvert}\int_{C_0} \sum\limits_{j=1}^M\frac{\partial(\eta_j ^*e^{-\imath2\bm{k}_j \cdot \bm{r}}) }{\partial t} \,{\rm d}\bm{r}\thickapprox\frac{\partial \eta_j}{\partial t}.
\end{align}
According to the approximation in \eqref{QDRG}, only the term containing $\eta_j$ yields a non-zero contribution because the other terms vanish when multiplied by a periodic function.

Performing the same operation on the right-hand side of \eqref{eqn:pfcdynamics} leads to
\begin{equation}
	\frac{\partial \eta_j}{\partial t} = \mathcal{L}_j\frac{\delta F}{\delta\eta_j^*},
	\label{eqn:apfc origin}
\end{equation}
where $\mathcal{L}_j = \nabla^2 + 2\imath\bm{k}_j\cdot\nabla - \lvert\bm{k}_j\rvert^2$. 

\begin{remark}
	\label{remark2}
	To simplify the calculation, we employ a long-wavelength limit~\cite{ElderAmplitude2010}, which allows us to ignore the high-order derivative terms $\nabla^2$ and $2\imath\bm{k}_j\cdot\nabla$ in $\mathcal{L}_j$. Specifically, as the wave number $\bm{k} \rightarrow 0$, the following result holds:
	\begin{equation}
		\label{eqn:remark2}
		\lim_{\bm{k}\rightarrow 0 } \frac{1}{\varepsilon^d}\int_{C_{\varepsilon}} ( \nabla^2 + 2\imath\bm{k}_j\cdot\nabla )\widehat{\eta}_je^{\imath\bm{k}\cdot\frac{\bm{x}}{\varepsilon}} \,{\rm d}\bm{x} = 0,
	\end{equation}
	where $\widehat{\eta}_j$ denote the Fourier coefficient of $\eta_j$. For the simplicity of presentation, the proof of~\eqref{eqn:remark2} is provided in Appendix~\ref{apd:sub:proofs}.
\end{remark}

Based on Remark~\ref{remark2}, we approximate $\mathcal{L}_j\thickapprox -\lvert\bm{k}_j\rvert^2$. Hence, the dynamic equation of the APFC model~\eqref{eqn:apfc origin} becomes
\begin{equation}
	\frac{\partial \eta_j}{\partial t} = -\lvert\bm{k}_j\rvert^2 \frac{\delta F}{\delta \eta_j^*},
	\label{eqn:apfcdynamics}
\end{equation}
where
\begin{equation}
	\frac{\delta F}{\delta \eta_j^*}
	=\big[ B_0 + B_x(\nabla^2 + 2\imath\bm{k}_j\cdot\nabla)^2 + 3v(A^2 - \lvert\eta_j\rvert^2) \big] \eta_j + \frac{\delta f^s}{\delta \eta_j^*}.
	\label{eqn:apfcdynamics-rightterm}
\end{equation}
The corresponding free energy functional with respect to $\eta_j$ is given by
\begin{equation}
	F = \frac{1}{\lvert\Omega\rvert}\int_{\Omega} \Big[\frac{B_0}{2}A^2 + \frac{3v}{4}A^4 + \sum_{j=1}^M \big(B_x\lvert\varsigma_j \eta_j\rvert^2   - \frac{3v}{2} \lvert\eta_j\rvert^4\big) + f^s({\eta_j},{\eta_j^*}   )  \Big] \,{\rm d}\bm{r},
	\label{eqn:energyfunctionalineta}
\end{equation}	
where $\varsigma_j := \nabla^2 + 2\imath\bm{k}_j\cdot \nabla $. The derivation details of~\eqref{eqn:apfcdynamics-rightterm} and~\eqref{eqn:energyfunctionalineta} are given in the Appendix~\ref{sec:sub:apd}.

The term $A^2$ in~\eqref{eqn:apfcdynamics-rightterm} is defined by
\begin{equation}
	A^2 := 2 \sum_{j=1}^M \lvert\eta_j\rvert^2.
	\label{eqn:A2}
\end{equation}
This parameter remains constant within solid and ordered phases but decreases in the presence of defects and interfaces. The functional $ f^s(\eta_j, \eta_j^*)$ corresponds to complex polynomials of $ \eta_j $ and $ \eta_j^* $, aligned with the appropriate symmetry.

\paragraph{Analyticity of solutions.}

The regularity of solutions has rarely been addressed in previous studies on phase-field crystal models. In this work, we provide a brief discussion of the analyticity of solutions to~\eqref{eqn:apfcdynamics}. At the steady state, the system reduces to a nonlinear elliptic system, for which Morrey~\cite{1958Morrey} established fundamental results ensuring analyticity of solutions. On the other hand, for the time-dependent case, the Cauchy–Kovalevskaya theorem~\cite{evans2022partial} applies: for parabolic PDE systems with analytic coefficients and analytic, non-characteristic initial data, the solution is analytic over a short time interval. Combining these observations, we obtain the following result.

\begin{theorem}
\label{thm:analytic}
For the nonlinear parabolic PDE system~\eqref{eqn:apfcdynamics}, 
if the coefficients are analytic and the initial data are analytic and non-characteristic, 
then the solution $\eta_j$ is real analytic in space and time, at least locally in time.
\end{theorem}

\begin{remark}
\label{remark:analyticity}
The analyticity established in Theorem~\ref{thm:analytic} provides partial justification for applying Fourier pseudospectral methods to the APFC model. To our knowledge, this work presents the first explicit statement of solution analyticity for the APFC system. While Morrey~\cite{1958Morrey} addressed the steady-state case in elliptic settings, and short-time analyticity for the full parabolic system follows from the Cauchy–Kovalevskaya theorem~\cite{evans2022partial}, long-time analyticity remains an open problem. This persistence may arise from the combination of short-time analyticity and energy dissipation. The numerical experiments in Section~\ref{sec:numerical experiments} further support this regularity property, demonstrating convergence rates consistent with analytic solutions.
\end{remark}

\begin{assumption}
\label{assumption}
For the nonlinear parabolic PDE system~\eqref{eqn:apfcdynamics}, 
if the coefficients are analytic and the initial data are analytic and non-characteristic, the solutions $\eta_j(\bm{x},t)$ remain real analytic in space and time for any $t>0$.
\end{assumption}

In the APFC model \eqref{eqn:apfcdynamics}, the crystal lattice symmetry is determined by selecting a suitable set of $M$ vectors $\{\bm{k}_j\}_{j=1}^{M}$. Below, we present two examples of lattice configurations considered in this work.
\begin{example}\label{ex}
	The reciprocal-space vectors used to characterize the crystal symmetry and $ f^s $ are determined by the chosen lattice symmetry. For simplicity, we present the results for two types of lattices illustrated in Figure~\ref{fig:structure of lattice}, with computational details provided in Appendix~\ref{sec:sub:apd}. 
	
	\begin{figure}[htbp]
		\centering
		\subfloat[Triangular lattice]{\includegraphics[width=.4\linewidth]{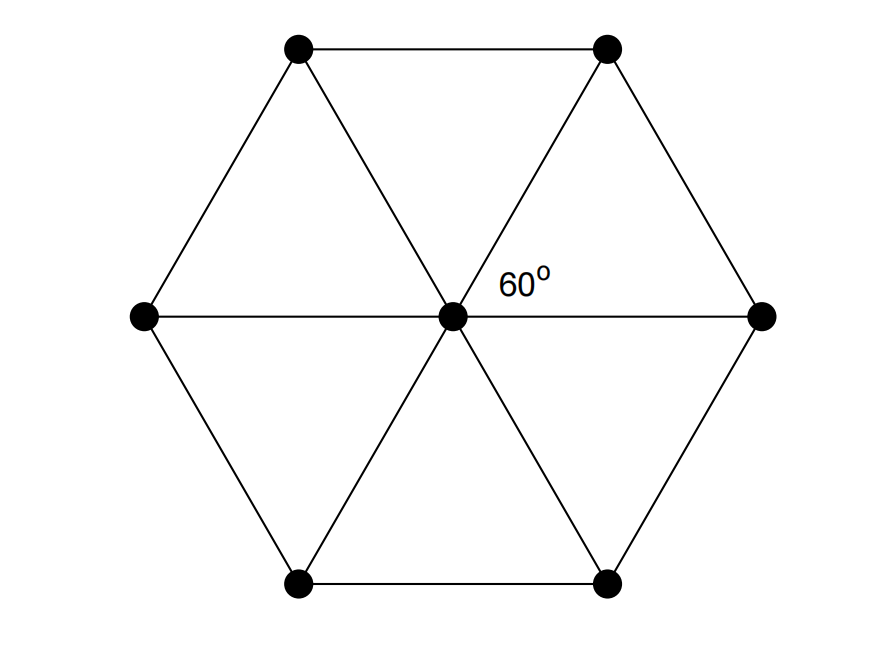}}~~
		\subfloat[BCC lattice]{\includegraphics[width=.4\linewidth]{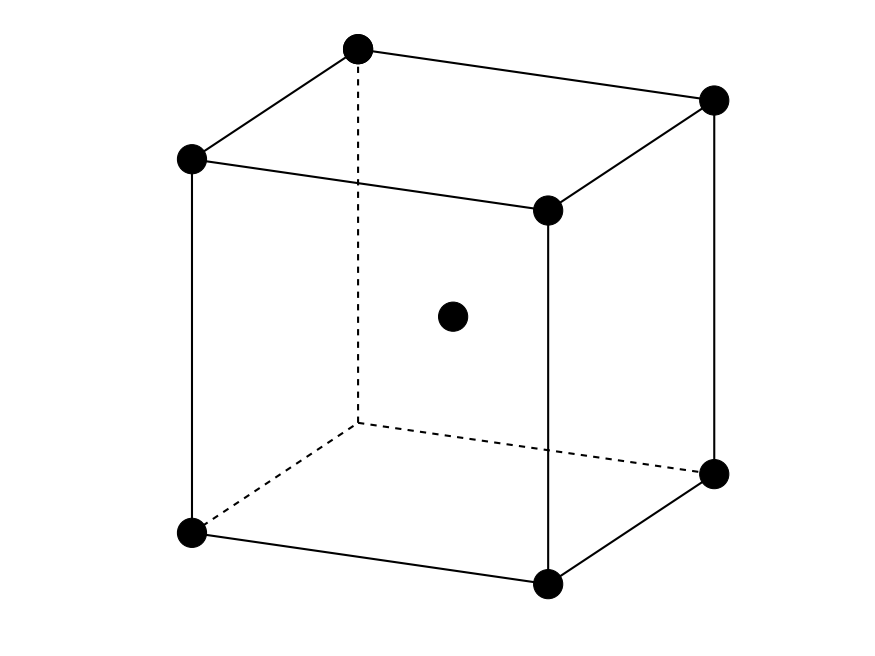}}~~
		\caption{The illustration of unit structures of triangular lattice and BCC lattice.}
		\label{fig:structure of lattice}
	\end{figure}
	
	\begin{itemize}
		\item For triangular lattice with $M=3$, the complex polynomial functional $f^{s}$ in \eqref{eqn:apfcdynamics-rightterm} reads
		\begin{align*}
			f^{s} = f^{s}_{\rm{tri}} &= -2\gamma(\eta_1\eta_2\eta_3 + 
			{\rm c.c.}),
		\end{align*}
    while the derivatives are given by
		\begin{align*}
			\frac{\delta f^{s}_{\rm tri}}{\delta \eta_j^*} &= -2\gamma\prod_{i\not=j}^{3}\eta_i^{*}, \qquad j = 1,2,3.
		\end{align*}
		
		\item For body-centered cubic (BCC) lattice with $M=6$, the complex polynomial functional $f^{s}$ in \eqref{eqn:apfcdynamics-rightterm} reads
		\begin{equation*}
			\begin{aligned}
				f^{s} = f^{s}_{\rm bcc} = -&2\gamma(\eta_1^*\eta_2\eta_3 + \eta_2^*\eta_3\eta_5 + \eta_3^*\eta_1\eta_6 + \eta_4^*\eta_5^*\eta_6^* + {\rm c.c.})\\ 
				&+ 6v(\eta_1\eta_3^*\eta_4^*\eta_5^* + \eta_2\eta_1^*\eta_5^*\eta_6^* +\eta_3\eta_2^*\eta_6^*\eta_4^* + {\rm c.c.}), 
			\end{aligned}
		\end{equation*}
		while the derivatives are given by 
		\begin{align*}
			\frac{\delta f^s_{\rm bcc}}{\delta \eta_i^*} &= -2\gamma(\eta_k\eta_n^* + \eta_j\eta_l) + 6v(\eta_k\eta_l\eta_m + \eta_j\eta_m^*\eta_n^*), \\ 
			\frac{\delta f^s_{\rm bcc}}{\delta \eta_l^*} &= -2\gamma(\eta_m^*\eta_n^* + \eta_i\eta_j^*) + 6v(\eta_i\eta_k^*\eta_m^* + \eta_k\eta_j^*\eta_n^*),
		\end{align*}
		where all the equations for the amplitudes are obtained by permutations on the groups $(i,j,k)=(1,2,3)$ and $(l,m,n)=(4,5,6)$.
	\end{itemize}
	
\end{example}

\section{Numerical Method for the APFC Model}
\label{sec:numerical methods}

Given the lattice symmetry and periodic boundary conditions, the Fourier pseudospectral method in the spatial direction is well-suited for nonlinear and high-order PDEs due to its high accuracy and straightforward implementation. In Section~\ref{sec: Fourier pseudospectral method}, we introduce the basic setup for the Fourier pseudospectral method. Then, in Section~\ref{sec: Implementation of Fourier pseudospectral method}, we provide a detailed explanation of its principles and practical implementation for solving the APFC model~\eqref{eqn:apfcdynamics} using the Fourier pseudospectral method.

\subsection{Basic setup}
\label{sec: Fourier pseudospectral method}

In pseudospectral methods, PDEs are solved at collocation points in physical space, where spatial derivatives are computed using orthogonal functions like trigonometric and Chebyshev polynomials. These functions are evaluated using techniques such as matrix multiplication, fast Fourier transform (FFT), or convolution. The Fourier pseudospectral method used in this study follows the procedure outlined in \cite{jiang2024numerical}, which includes selecting collocation points, discretizing the test function by computing the inner product of the equation, obtaining Fourier coefficients, and performing an inverse Fourier transform to obtain the desired function \cite{2011Spectral}.

To that end, given an even integer $N\in \mathbb{N}_0$, we denote 
\begin{equation*}
	K_N^d := \left\{\bm{k} = (k_j)_{j=1}^{d}\in \mathbb{Z}^d: -\frac{N}{2}\leq k_j < \frac{N}{2}\right\},
\end{equation*}
where $d$ represents dimension. Let $\mathbb{T}^d := (\mathbb{R}/2\pi\mathbb{Z})^d$ be the $d$-dimensional torus. We discretize $\mathbb{T}^d$ using the fundamental domain $[0, 2\pi)^d$ with uniform spatial discretization parameter $h = 2\pi/N$. The grid $\mathbb{T}^d_N$ consists of points $\bm{r}_p = (x_{1,l_1}, x_{2,l_2}, \ldots, x_{d,l_d})$, where $x_{i,l_i} = l_ih$, for $0 \leq l_i < N$, $l_i \in \mathbb{N}_0$.

We define the grid function space
\begin{equation*}
	\mathcal{G}_N:=\big\{u:\mathbb{T}_N^d\mapsto \mathbb{C} ~\lvert~ u\text{ is } \mathbb{T}^d_N-\text{periodic}\big\}.
\end{equation*}
For any periodic grid functions $u, v\in\mathcal{G}_N$, the $\ell^2$-inner product is defined as
\begin{equation*}
	\langle u,v\rangle_N := \frac{1}{(2\pi N)^d}\sum_{\bm{r}_p\in\mathbb{T}^d_N}u(\bm{r}_p)v^{*}(\bm{r}_p).
\end{equation*}
For $\bm{k}, \bm{l}\in\mathbb{Z}^d$, we have the discrete orthogonality condition
\begin{equation}
\langle e^{\imath\bm{k}\cdot\bm{r}_p},e^{\imath\bm{l}\cdot\bm{r}_p}\rangle_N = \sum_{\bm{r}_p\in\mathbb{T}^d_N}e^{\imath(\bm{k}-\bm{l})\cdot\bm{r}_p} =\left\{
\begin{aligned}
    1,\quad&\bm{k} = \bm{l} + N\bm{m},\text{ }  \bm{m}\in\mathbb{Z}^d\\
    0,\quad&{\rm otherwise}\\
\end{aligned}
\right..
\label{eqn:orthogonality}
\end{equation}
The discrete Fourier coefficient of $u\in\mathcal{G}_N$ is given by
\begin{equation}
\label{eq:disFourierCoeff}
u_{\bm{k}} = \langle u,e^{\imath\bm{k}\cdot\bm{r}_p}\rangle_N,\quad \bm{k}\in K_N^d.
\end{equation}

We can also perform the discrete Fourier transform of the periodic function $u$ by 
\begin{equation}
	u(\bm{r}_p) = \sum_{\bm{k}\in K^d_N}u_{\bm{k}}e^{\imath\bm{k}\cdot\bm{r}_p}, \quad \bm{r}_p\in \mathbb{T}_N^d.
	\label{eqn:discrete fourier transform}
\end{equation}
The trigonometric interpolation of the periodic function is
\begin{equation}
\label{eq:IN}
I_Nu(\bm{r}) = \sum_{\bm{k}\in K^d_N}u_{\bm{k}}e^{\imath\bm{k}\cdot\bm{r}}.
\end{equation}
It is straightforward to see that $I_Nu(\bm{r}_p) = u(\bm{r}_p)$. In practice, one can use the $d$-dimensional FFT to obtain the Fourier coefficients $u_{\bm{k}}$ in \eqref{eqn:discrete fourier transform}.

\begin{remark}
In addition to the Fourier spectral method discussed above, the finite element method is also commonly used for solving the APFC model~\cite{2017Controlling}. However, it demands more computational resources and requires careful consideration of spatial discretization. For the sake of completeness, we include the implementation details of finite element method in solving the APFC model in Appendix~\ref{sec:sub:fem}.
\end{remark}

\subsection{Implementation of Fourier pseudospectral method} 
\label{sec: Implementation of Fourier pseudospectral method}

In this section, we present a detailed implementation of solving the APFC model~\eqref{eqn:apfcdynamics} by employing Fourier pseudospectral method.

We begin by expanding the amplitude functions $\eta_j$ to

\begin{equation}
	\eta_j(\bm{r},t) = \phi_je^{\imath \bm{k}_j \cdot \bm{u}(\bm{r})}, \quad j = 1,\ldots,M,\label{amplitude}
\end{equation}
where $\bm{u}(\bm{r})$ corresponds to the displacement field with respect to the reference lattice~\cite{2017Controlling}. The equations of each $\eta_j$ are complex functions, and they are independent of each other. Substituting \eqref{amplitude} into \eqref{eqn:apfcdensity}, we can obtain
\begin{align}
	n(\bm{r},t) &= n_0(\bm{r},t) + \sum_{j=1}^M \big[\phi_j e^{\imath\bm{k}_j \cdot \bm{r}}e^{\imath\bm{k}_j \cdot \bm{u}(\bm{r})} + \phi_j e^{ -\imath\bm{k}_j \cdot \bm{r} }e^{-\imath\bm{k}_j \cdot \bm{u}(\bm{r})} \big]\\
	&=n_0(\bm{r},t) + \sum_{j=1}^M \big[\phi_j e^{\imath\bm{k}_j \cdot (\bm{r} + \bm{u}(\bm{r}))} + \phi_j e^{ -\imath\bm{k}_j \cdot (\bm{r} + \bm{u}(\bm{r}))}\big].
\end{align}
Compared with the density defined by~\eqref{eqn:apfcdensity}, we observe that $\bm{u}(\bm{r})$ represents the perturbation term, while $\phi_j e^{\imath\bm{k}_j\cdot\bm{r}}$ corresponds to the reference lattice. Here, $\phi_j$ represents the real value of the amplitude in the relaxed crystal. These quantities can be computed by minimizing the energy functional in \eqref{eqn:energyfunctional}, which is the equilibrium lattice state of the system. 

In our work, square and cube are employed as computational domains for two-dimensional and three-dimensional cases, respectively. Let $L$ be the side length of the computational domain. The grid points are $\bm{r}_p = (\frac{m_1L}{N}, \frac{m_2L}{N})\in\mathbb{T}_N^2$, $0\leq m_1,m_2<N$, and $\bm{r}_p = (\frac{m_1L}{N}, \frac{m_2L}{N}, \frac{m_3L}{N})\in\mathbb{T}_N^3$, $0\leq m_1,m_2,m_3<N$, $m_i\in\mathbb{N}_0$. The dimensions of $K_N^d$ in different situations are
\begin{equation}
	\dim(K_N^d) = N^d=:D,\quad d = 2, 3. 
	\label{dim-D}
\end{equation}

The amplitude functions $\eta_j$ are periodic on the Bravais lattice. Based on the relationship between the Bravais lattice and the reciprocal lattice, as well as their mathematical descriptions provided in Appendix~\ref{sec:Reciprocal Lattice}, we can derive their discrete Fourier transform from \eqref{eqn:periodic expansion} as follows:
\begin{equation*}
	\eta_j(\bm{r}_p,t)\approx I_N^d \eta_j(\bm{r}_p,t) = \sum_{\bm{k}\in K_N^d} \eta_{j_{\bm{k}}}(t)e^{\imath(B\bm{k})\cdot\bm{r}_p},\quad p = 0,1,\ldots, N^d-1,
\end{equation*}
where $\eta_{j_{\bm{k}}}(t) = \langle\eta_j, e^{\imath\bm{k}\cdot \bm{r}_p}\rangle_N.$ The first-order and second-order derivatives of $\eta_j$ in the Fourier pseudospectral method are given by

\begin{equation}
\label{eq:derivative}
\begin{aligned}
\nabla\eta_j &\approx \sum_{\bm{k}\in K_N^d}(B\bm{k})\eta_{j_{\bm{k}}}(t)e^{\imath(B\bm{k})\cdot \bm{r}_p},\\
\nabla^2\eta_j &\approx -\sum_{\bm{k}\in K_N^d}\lvert B\bm{k}\rvert^2\eta_{j_{\bm{k}}}(t)e^{\imath(B\bm{k})\cdot\bm{r}_p}.
\end{aligned}
\end{equation}

\subsubsection{Time discretization}

We now proceed to the discretization of the APFC dynamics. Substituting~\eqref{eq:derivative} into the APFC model~\eqref{eqn:apfcdynamics} yields
\begin{align*}
	&\sum_{\bm{k} \in K_N^d} \frac{\partial\eta_{j_{\bm{k}}}(t)}{\partial t}e^{\imath(B\bm{k})\cdot \bm{r}_p}\\  
	=&-\lvert \bm{k}_j\rvert^2 \left[ \sum_{\bm{k}\in K_N^d} \Big(B_0 + B_x\big(-(B\bm{k})^2-2\bm{k}_j\cdot({B\bm{k}})\big)^2\Big)\eta_{j_{\bm{k}}}(t)e^{\imath(B\bm{k})\cdot \bm{r}_p} + G_{\bm{k}}(\eta)\right],
\end{align*}
where
\begin{equation*}
	G(\eta) = G\big(t,\eta_j(\bm{r}_p,t)\big) := 3v(A^2- \lvert\eta_j\rvert^2)\eta_j + \frac{\delta f^s}{\delta \eta_j^*}
\end{equation*}
denotes the function depending on $\eta_1,\ldots,\eta_M, \eta_1^*, \ldots, \eta_M^*$; this function is treated explicitly in the time discretization.

Taking the inner product on both sides of the equation with the basis function $e^{\imath(B\bm{k})\cdot \bm{r}_q}$ and applying the orthogonality relation~\eqref{eqn:orthogonality}, we obtain
\begin{equation}
	\frac{\partial \eta_{j_{\bm{k}}}}{\partial t} = -\lvert\bm{k}_j\rvert^2 \big[B_0 + B_x\big(-(B\bm{k})^2-2\bm{k}_j\cdot(B\bm{k})\big)^2\big]\eta_{j_{\bm{k}}}(t) + G_{\bm{k}}(\eta).
	\label{eqn:convolution}
\end{equation}

For the time discretization, we employ a semi-implicit scheme, leading to
\begin{equation*}
	\frac{ \tilde{\eta}_{j_{\bm{k}}}^{(n+1)}- \tilde{\eta}_{j_{\bm{k}}}^{(n)}}{\Delta t} = -\lvert\bm{k}_j\rvert^2 \big[B_0 + B_x\big(-(B\bm{k})^2-2\bm{k}_j\cdot(B\bm{k})\big)^2\big]\tilde{\eta}_{j_{\bm{k}}}^{(n+1)}+ \tilde{G}_{\bm{k}}(\eta^{(n)}),
\end{equation*}
where $\Delta t$ is the time step size, and $\tilde{\eta}_{j_{\bm{k}}}^{(n)} =\tilde{\eta}_{j_{\bm{k}}}(t_n)$ is the numerical solution at time $t_n = n\Delta t$. The resulting iterative scheme is given by
\begin{equation}
\tilde{\eta}_{j_{\bm{k}}}^{(n+1)} = \frac{\tilde{\eta}_{j_{\bm{k}}} ^{(n)} - \Delta t\lvert\bm{k}_j\rvert^2\tilde{G}_{\bm{k}}(\eta^{(n)})}{1 + \Delta t\lvert\bm{k}_j\rvert^2\big[B_0 + B_x\big(-(B\bm{k})^2-2\bm{k}_j\cdot(B\bm{k})\big)^2\big] }.
\label{eqn:iterative format}
\end{equation}

The approximate solution $\tilde{\eta}_j(\bm{r}, t_n)$ is then reconstructed via
\begin{equation}
\tilde{\eta}_j (\bm{r}, t_n) = \sum_{\bm{k}\in K_N^d} \tilde{\eta}_{j_{\bm{k}}}(t_n)e^{\imath(B\bm{k})\cdot\bm{r}}.
\end{equation}

The procedure for calculating the displacement field $\bm{u}(\bm{r})$ from the numerical solutions $\tilde{\eta}_j$ is presented in Sections~\ref{sec:sub:strain_field_2D} and~\ref{sec:sub:strain_field_3D}.

Given the analyticity of $\eta_j$ established in Assumption~\ref{assumption} (see the discussion in Remark~\ref{remark:analyticity}), the Fourier pseudospectral method achieves spectral accuracy~\cite{2011Spectral}. We now present the corresponding interpolation error estimate in the $L^2$-norm.

\begin{theorem}
\label{thm:error-estimate}
Under the analyticity Assumption~\ref{assumption}, let $T>0$ be sufficiently large and $\eta_j$ denote the solution of~\eqref{eqn:apfcdynamics}. There exist positive constants \( C_{\rm s} \) and \( c \), independent of $\eta_j$ and $N$, such that
\[
\|I_N \eta_j(\cdot, T) - \eta_j(\cdot, T)\|_{L^2} \leq C_{\rm s} e^{-cN},
\]
where $I_N$ is the interpolation operator defined in~\eqref{eq:IN}.
\end{theorem}

\begin{remark}
\label{remark:error-estimate}
To rigorously establish the convergence of the fully discrete numerical solution $\tilde{\eta}_j(\cdot, t_n)$ to the exact solution $\eta_j(\cdot, t_n)$, a comprehensive error estimate that accounts for both spatial and temporal discretizations is required. Such a complete analysis typically hinges on a detailed investigation of the scheme's energy dissipation properties. While our numerical experiments in Section~4 clearly demonstrate exponential decay over long time intervals, a full mathematical justification of this behavior lies beyond the scope of the present paper and is therefore deferred to future work.
\end{remark}

We now analyze the computational complexity of solving the iterative scheme \eqref{eqn:iterative format} at each time step. This formulation can be efficiently implemented using the Fourier pseudospectral method in $d$-dimensional reciprocal space. While the linear terms are handled directly, the nonlinear contributions $G_{\bm{k}}(\eta)$ in \eqref{eqn:convolution} involve convolution operations that would be computationally expensive if evaluated directly in Fourier space. However, these nonlinear terms can be efficiently computed as local dot products in physical space by leveraging the Fast Fourier Transform (FFT) algorithm~\cite{zhang2008efficient}. Consequently, the overall complexity per time step is reduced to $\mathcal{O}(D \log D)$, where $D$ denotes the number of degrees of freedom defined in \eqref{dim-D}. This approach achieves high numerical efficiency and significantly accelerates computations, as will be demonstrated in the following section.

\section{Numerical Experiments}
\label{sec:numerical experiments} 

In this section, we conduct numerical experiments on both a two-dimensional triangular lattice and a three-dimensional body-centered cubic (BCC) lattice. We introduce quadrupole edge dislocations into the lattice structure with periodic boundary conditions and compute the corresponding variables (e.g., $A^2$ defined in \eqref{eqn:A2}) and the strain field. To further validate the proposed numerical methods, we compare the results with those obtained from continuum elasticity theory of dislocations~\cite{Cai2006A, olson2023elastic}.
\subsection{Two-dimensional triangular lattice}
\label{sec:numericalexperiments:2d}

We first consider a two-dimensional triangular lattice described in Example~\ref{ex}. Let $\phi_j = \phi_0$ represent the amplitude of the equilibrium crystal. By minimizing the energy functional given in \eqref{eqn:energyfunctionalineta}, we obtain $\phi_0$ as follows:
\begin{equation*}
	\phi_0 = \frac{\gamma + \sqrt{\gamma^2 - 15v B_0}}{15v}.
\end{equation*}
Further computational details can be found in Appendix~\ref{sec:sub:phi0}. A square computational domain $L_x \times L_y$, with edge lengths $L_x = L_y = 80\pi$, is utilized for numerical simulations. To simulate crystal dislocations, only horizontal shear stress parallel to the $x$-axis is applied, with no force exerted in the $y$-direction. The initial condition $\eta_{j,0}$ of the amplitude functions can be determined based on $\eta_j = \phi_0 e^{\imath \bm{k}_j \cdot \bm{u}(\bm{r})}$ by defining $\bm{u}(\bm{r})$ as the displacement of the lattice. Given $\bm{r} := x \widehat{\bm{x}} + y \widehat{\bm{y}}$ as the spatial position vector, the displacement $\bm{u}(\bm{r})$ is given by
\begin{equation}
	\bm{u}(\bm{r}) = \left\{
	\begin{aligned}
		&-u_x \widehat{\bm{x}}, \quad &\frac{L_y}{4} < y < \frac{3L_y}{4} \\
		&+u_x \widehat{\bm{x}}, \quad &\text{elsewhere}
	\end{aligned}
	\right.,
\end{equation}
where $u_x = (a_x/L_x)x$ with $a_x = 4\pi/\sqrt{3}$ and $u_y = 0$. With this choice, a strain of $\varepsilon = \pm a_x/L_x$ is applied to the corresponding region. 

We solve the APFC model~\eqref{eqn:apfcdynamics} using the Fourier pseudospectral method. To validate the accuracy of the numerical scheme described in Section~\ref{sec:numerical methods}, a reference solution $\eta_j^{*}(\bm{r},T)$ is computed with a sufficiently small time step $\tau = 1 \times 10^{-4}$, a fine spatial mesh size $h = \pi /128$, and a final time $T=4$. The choice of $T$ is determined empirically so that the system attains a steady state, as confirmed by numerical experiments.

\begin{figure}[htbp]
	\centering
	\includegraphics[width=.5\linewidth]{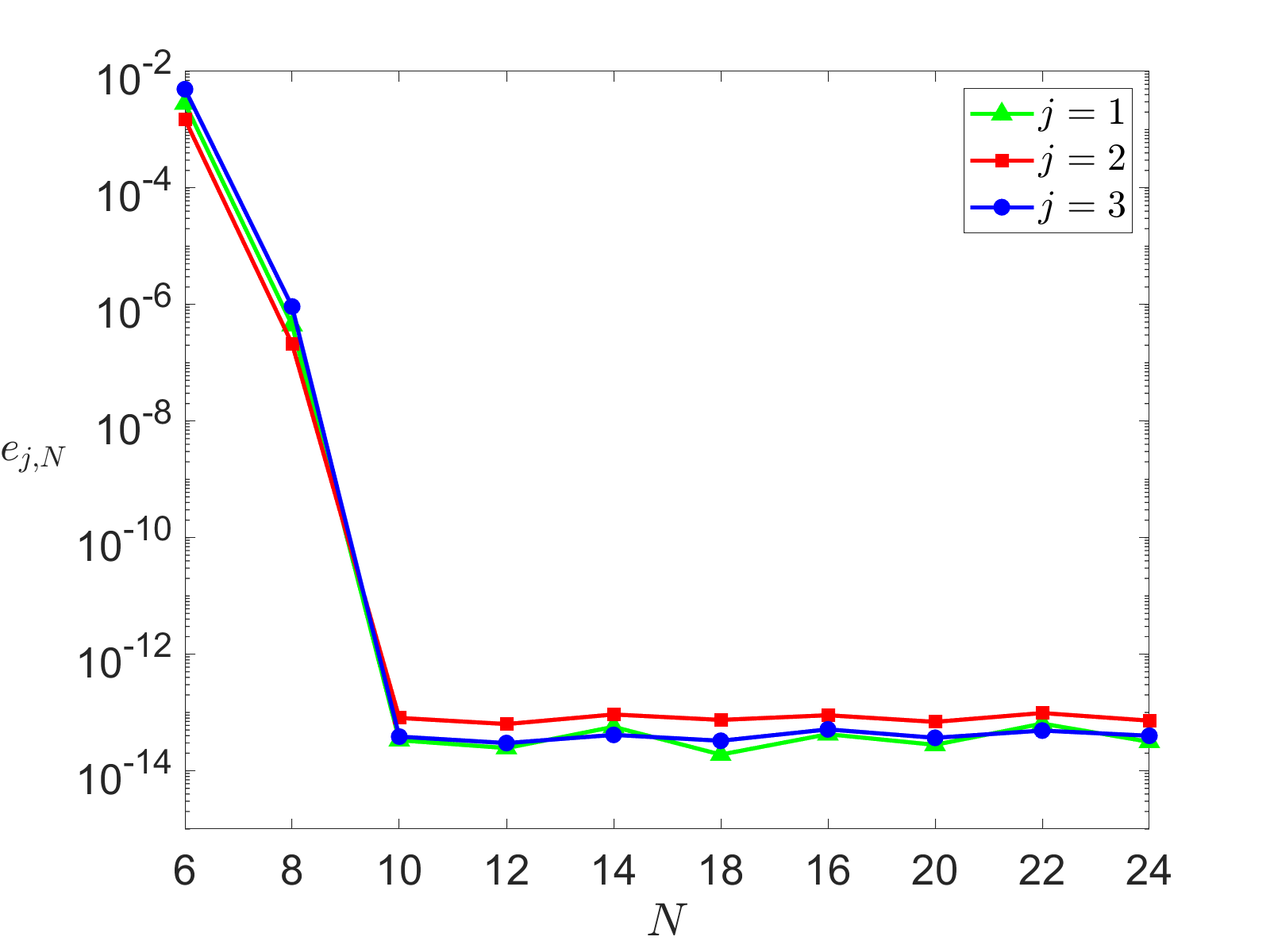}
	\caption{Triangular lattice: numerical error $e_{j,N}$ for $j=1,2,3$, $d=2$ as a function of $N$.}
	\label{fig:error-2D}
\end{figure}

Let the reference solution of~\eqref{eqn:apfcdynamics} be expressed as
\begin{equation}
\label{eq:exact}
\eta_j^{*}(\bm{r},T) = \sum_{\bm k\in K_{N}^d}\eta_{j_{\bm{k}}}^{*} e^{\imath (B\bm{k})\cdot \bm{r}}, \quad j=1,2,3,
\end{equation}
and define the numerical error by
\begin{equation}
\label{eq:error}
e_{j,N} = \|\eta_j^*(\cdot,T) - \tilde{\eta}_j(\cdot,T)\|_{L^2}^2
= \sum_{\bm k\in K_N^d} \big| \eta_{j_{\bm{k}}}^*(T) - \tilde{\eta}_{j_{\bm{k}}}(T) \big|^2.    
\end{equation}
Hence, the error of the Fourier pseudospectral method is measured by the convergence of the Fourier coefficients. Figure~\ref{fig:error-2D} shows the error $e_{j,N}$ for $j=1,2,3$ with $d=2$ as a function of $N$. For the nonlinear system under the assumptions of Theorem~\ref{thm:analytic}, $\eta_j$ is analytic, and the results confirm spectral accuracy, in agreement with Theorem~\ref{thm:error-estimate}.

To further demonstrate the computational efficiency, we report the CPU time (in seconds) required to solve the model for various values of $N$, which correspond to the degrees of freedom in each spatial direction (see Section~\ref{sec: Fourier pseudospectral method}). The time step size, $\Delta t$, is set to 0.1. Table~\ref{tab:2Dtime} shows the time required for different values of $N$. The CPU time scales approximately as $O(N^{1.16})$, demonstrating the high efficiency of our model solution process. This efficiency is further underscored by the relatively low computational effort required, as discussed in Section~\ref{sec: Implementation of Fourier pseudospectral method}.
\begin{table}[htbp]
	\caption{Triangular lattice: CPU time (s) vs. $N$.}
	\label{tab:2Dtime}       
	\begin{tabular}{lllllll}
		\hline\noalign{\smallskip}
		$N$&32&64&128&256&512&1024 \\
		\noalign{\smallskip}\hline\noalign{\smallskip}
		CPU time (s) &	0.180	&	0.462&	1.042&	1.766&	3.320&	13.538\\
		\noalign{\smallskip}\hline
	\end{tabular}
\end{table}

\subsubsection{Density $n$ and amplitude parameter $A^2$}

\begin{figure}[htbp]
	\centering
	\subfloat[$n$ of the perfect lattice]{\label{fig:A2-2D-a}
		\includegraphics[width=.24\linewidth]{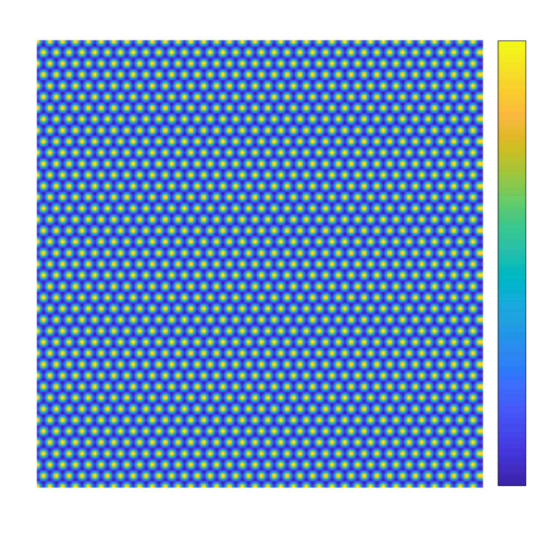}}~~
	\subfloat[$n$ of the defect lattice]{\label{fig:A2-2D-b}\includegraphics[width=.42\linewidth]{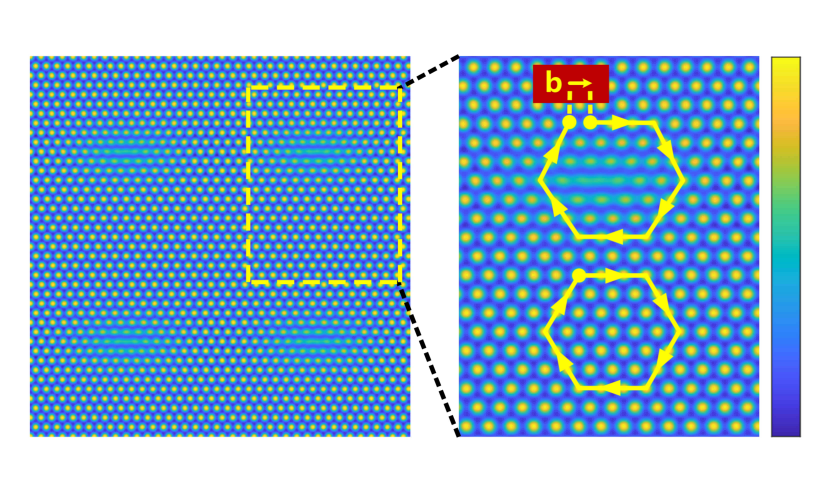}}~~
	\hspace{-4mm}
	\subfloat[$A^2$]{\label{fig:A2-2D-c}\includegraphics[width=.32\linewidth]{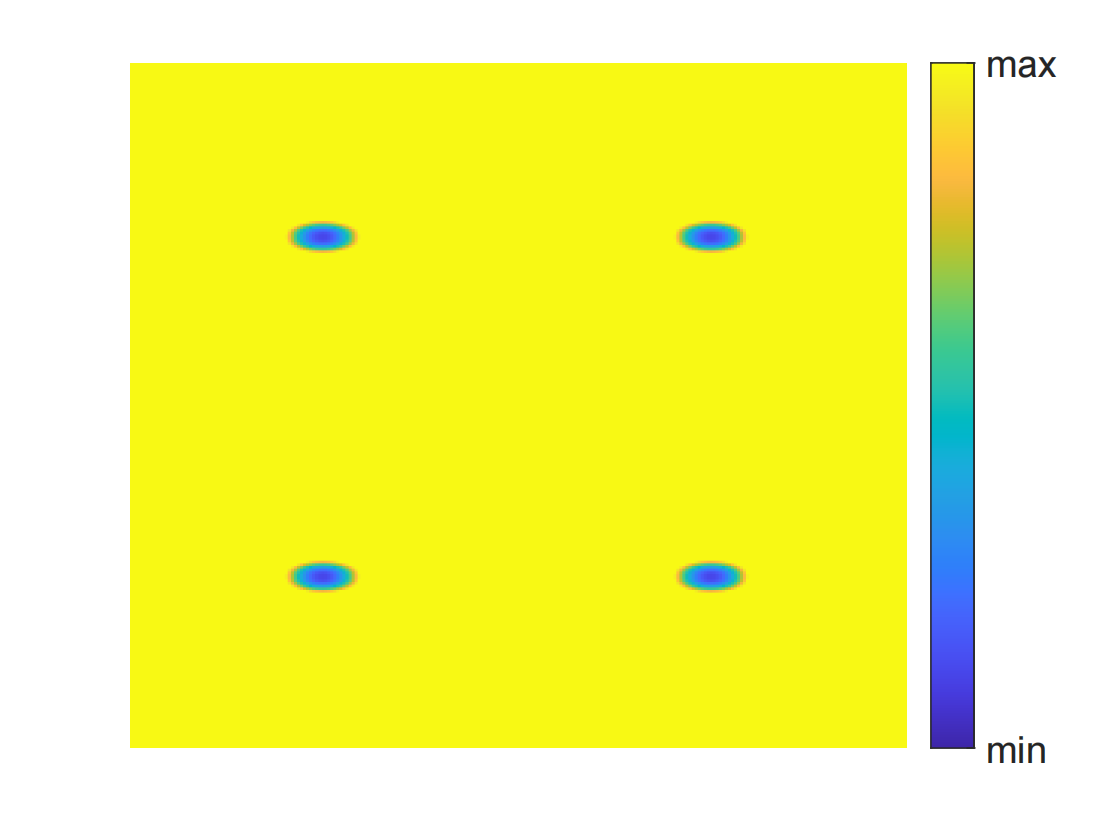}}
	\caption{Triangular lattice: The illustration of the density $n$ defined in \eqref{eqn:apfcdensity} of the perfect lattice and defect lattice, and the parameter $A^2$ defined in \eqref{eqn:A2}. Both are colored by values.}
	\label{fig:A2-2D}
\end{figure}

By solving APFC model \eqref{eqn:apfcdynamics},  we can obtain the amplitude functions $\eta_j$. These functions can then be used to reconstruct density $n$ by \eqref{eqn:apfcdensity} and calculate $A^2$ by \eqref{eqn:A2}. Figure~\ref{fig:A2-2D} presents the visualization of the density $n$ defined in \eqref{eqn:apfcdensity} and the parameter $A^2$ defined in \eqref{eqn:A2}, using the parameters $B_0 = 0.02$, $B_x = 0.98$, $v = 1/3$, and $\gamma = 1/3$ from \eqref{eqn:energyfunctionalineta} as provided in~\cite{2017Controlling}. Figure~\ref{fig:A2-2D-a} and Figure~\ref{fig:A2-2D-b} display the density $n$ of the perfect lattice and the defect lattice, which can be reconstructed from \eqref{eqn:apfcdensity}. In Figure~\ref{fig:A2-2D-b}, the area near the defect is extracted, and the Burgers vector can be clearly seen. Figure~\ref{fig:A2-2D-c} depicts four blue ellipses representing the regions of defects formed as the system stabilizes, indicating the distribution of dislocations, specifically a quadrupole dislocation. This numerical distribution satisfies the property of the parameter $A^2$: it remains constant within ordered phases but decreases in the presence of defects. 

Thus, the APFC model effectively retains atomistic features within a continuum framework, functioning as a mesoscale model that connects atomistic and continuum approaches.

\subsubsection{Strain field and comparison with continuum elasticity}
\label{sec:sub:strain_field_2D}

To demonstrate the accuracy of the APFC model solutions, we calculate the strain field and compare with continuum elastic theory. 

\paragraph{Strain field.} 

For small deformations, the strain tensor $\bm{\varepsilon}$ can be expressed as 
\begin{equation}
	\bm{\varepsilon} = \frac{1}{2}\left[(\nabla {\bm u}) + (\nabla {\bm u})^T\right].
	\label{eqn:strain}
\end{equation}
By solving the APFC model \eqref{eqn:apfcdynamics}, we can obtain $\eta_j$. The relationship between $\eta_j$ and ${\bm u}$ is established by introducing an intermediate variable:
\begin{equation}
	\varphi_j = \bm{k}_j \cdot \bm{u}(\bm{r}), \quad j=1,\ldots,M.
	\label{N-eqs}
\end{equation}
Thus, the equation \eqref{amplitude} can be written as $\eta_j = \phi_0 e^{\imath \varphi_j}$. Furthermore, $\varphi_j$ can be expressed in terms of $\eta_j$ as follows:  
\begin{equation}
	\varphi_j = \arg(\eta_j) = \arctan\left(\frac{\mathrm{Im}(\eta_j)}{\mathrm{Re}(\eta_j)}\right),
	\label{varphi by eta}
\end{equation}
where \eqref{N-eqs} defines a system of $M$ linear equations related to $u$. We employ the least squares method to solve this system. After some straightforward calculations, the least squares solution of \eqref{N-eqs} in the sense of the Moore-Penrose inverse is expressed as follows:
\begin{equation}
	\bm{u}(\bm{r}) = \left\{
	\begin{array}{lr}
		u_x = -\frac{\sqrt{3}}{3}\varphi_1 + \frac{\sqrt{3}}{3}\varphi_3, \\
		u_y = -\frac{1}{3}\varphi_1 + \frac{2}{3}\varphi_2 - \frac{1}{3}\varphi_3.
	\end{array}
	\right.
	\label{u2D}
\end{equation}

To calculate the strain field according to \eqref{eqn:strain}, we need to obtain $\nabla {\bm u}$. To this end, based on the expression in \eqref{u2D}, we compute the partial derivatives of $\varphi_j$,
\begin{equation}
	\frac{\partial \varphi_j}{\partial \xi} = \frac{1}{\lvert\eta_j\rvert^2}\left[ \frac{\partial \mathrm{Im} (\eta_j)}{\partial \xi} \mathrm{Re}(\eta_j) -
	\frac{\partial \mathrm{Re} (\eta_j)}{\partial \xi} \mathrm{Im}(\eta_j)\right], \quad \xi = x,y.
	\label{derivation}
\end{equation}
Thus, the strain field can be computed solely from $\eta_j$, obtained by solving the APFC model \eqref{eqn:apfcdynamics}, without the need for additional data. The formulas for the components of the strain field are:
\begin{equation}
	\begin{split}
		\varepsilon_{xx} &= -\frac{\sqrt{3}}{3}\frac{\partial \varphi_1}{\partial x} + \frac{\sqrt{3}}{3}\frac{\partial \varphi_3}{\partial x}, \\
		\varepsilon_{yy} &= -\frac{1}{3}\frac{\partial \varphi_1}{\partial y} + \frac{2}{3}\frac{\partial \varphi_2}{\partial y} - \frac{1}{3}\frac{\partial \varphi_3}{\partial y}, \\
		\varepsilon_{xy} &= -\frac{\sqrt{3}}{3}\frac{\partial \varphi_1}{\partial y} + \frac{\sqrt{3}}{3}\frac{\partial \varphi_3}{\partial y} - \frac{1}{3}\frac{\partial \varphi_1}{\partial x} + \frac{2}{3}\frac{\partial \varphi_2}{\partial x} - \frac{1}{3}\frac{\partial \varphi_3}{\partial x}.
	\end{split}
\end{equation}

\begin{figure}[htbp]
	\centering
	\subfloat[$\varepsilon_{xx}$]{\includegraphics[width=.32\linewidth]{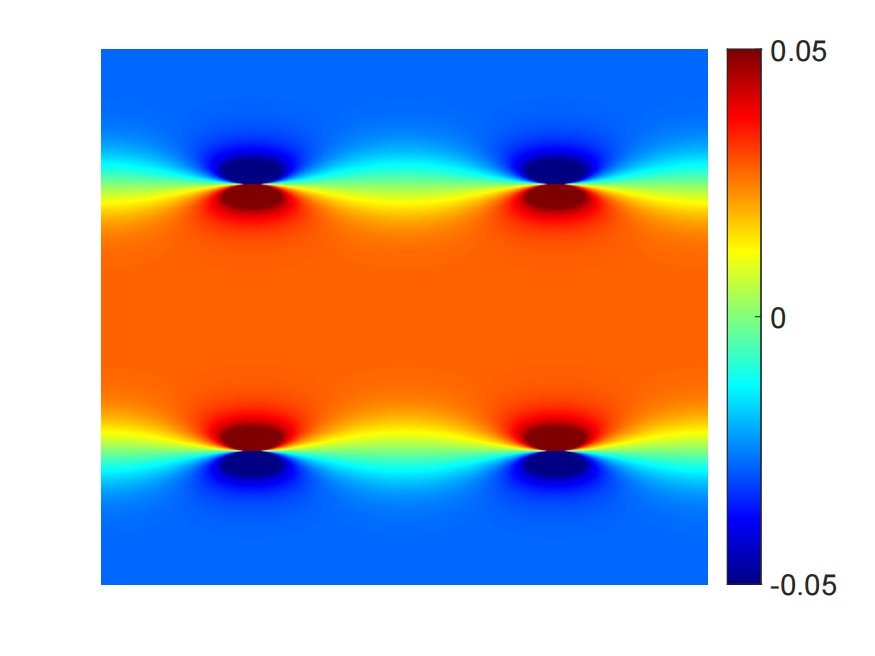}}~~
	\subfloat[$\varepsilon_{xy}$]{\includegraphics[width=.32\linewidth]{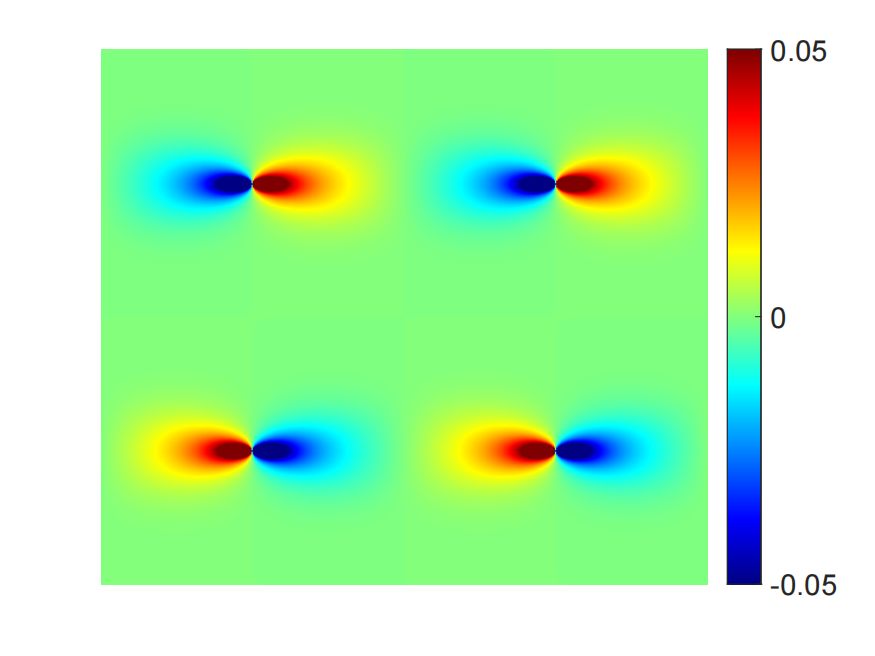}}~~
	\subfloat[$\varepsilon_{yy}$]{\includegraphics[width=.32\linewidth]{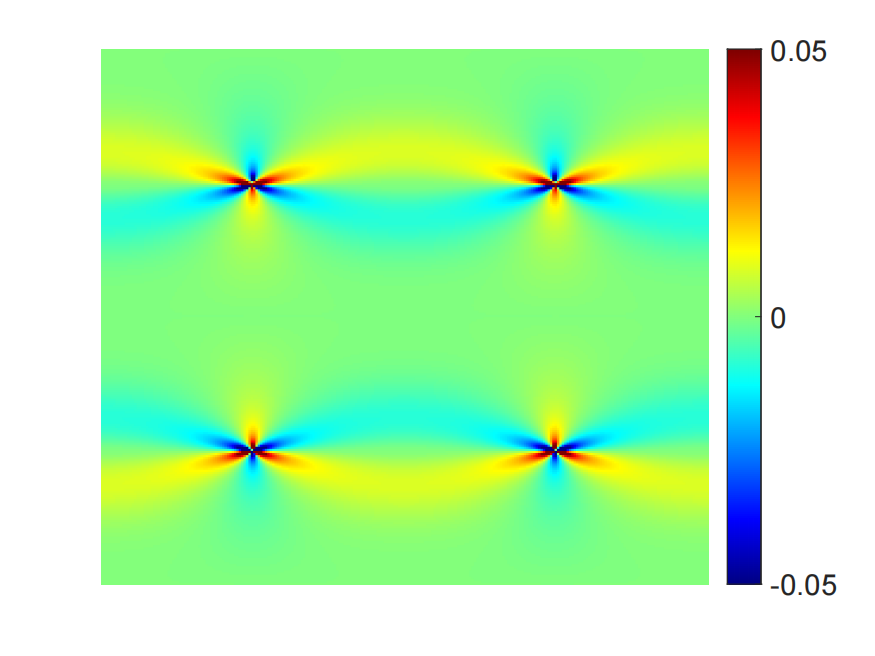}}
	\caption{The components of strain field for two-dimensional triangular lattice: (a) $\varepsilon_{xx}$, (b) $\varepsilon_{xy}$, and (c) $\varepsilon_{yy}$.}
	\label{fig:strainfielddislocation}
\end{figure}

Figure~\ref{fig:strainfielddislocation} illustrates the components of the strain field in a two-dimensional triangular lattice. The distribution of the strain field for a quadrupole dislocation is observed, and notably, the field distribution at each defect core aligns with theoretical predictions discussed in prior work~\cite{anderson2017theory}.

\paragraph{Comparison with continuum elasticity theory.}

We compare the APFC solution with continuum elasticity theory to describe the elastic field of a dislocation, assuming that the dislocation characteristics, such as the Burgers vector $\bm{b}$ and the elastic constants, are known~\cite{anderson2017theory}. Without loss of generality, we set the position of the dislocation core at $\bm{x}_c = (0,0)$ and that the Burgers vector $\bm{b}$ aligns with the $x$-axis. To mitigate the singularity inherent at the dislocation core, we adopt the regularization method proposed in~\cite{Cai2006A}. The stress field components can be computed as follows:

\begin{equation}
	\begin{split}
		\sigma_{xx} / \sigma_0 &= -y(3\zeta^2 + 3x^2 + y^2), \\
		\sigma_{yy} / \sigma_0 &= x(\zeta^2 + x^2 - y^2), \\
		\sigma_{yx} / \sigma_0 &= \sigma_{xy} / \sigma_0 = x(\zeta^2 + x^2 - y^2), 
	\end{split}
\end{equation}
where 
\begin{equation}
	\sigma_0 = \frac{E\lvert\bm{b}\rvert}{2\pi(1-\nu^2)(\zeta^2 + x^2 + y^2)^2},
\end{equation}
with $\zeta = \lvert\bm{b}\rvert/2$ representing the regularization term, $ \lvert\bm{b}\rvert = 4/\sqrt{3} $, and $E$ and $\nu$ representing Young’s modulus and Poisson’s ratio, respectively.

The strain field can be determined using Hooke’s law:
\begin{equation}
	\bm{\varepsilon}^{\rm CE} = \left[\frac{1+\nu}{E}\right]\bm{\sigma}^{\rm tot} - \left(\frac{\nu}{E}\right) {\rm tr}(\bm{\sigma}^{\rm tot})\bm{I},
\end{equation}
where $\bm{\sigma}^{\rm tot}$ is the total stress field. The strain field in each direction is obtained as follows:
\begin{equation}
	\begin{split}
		\varepsilon_{xx} &= \frac{1}{E}(\sigma_{xx} - \nu\sigma_{yy}), \\
		\varepsilon_{yy} &= \frac{1}{E}(\sigma_{yy} - \nu\sigma_{xx}), \\
		\varepsilon_{xy} &= \frac{1+\nu}{E}\sigma_{xy}.
	\end{split}
\end{equation}

\begin{figure}[htbp]
	\centering
	\includegraphics[width=.5\linewidth]{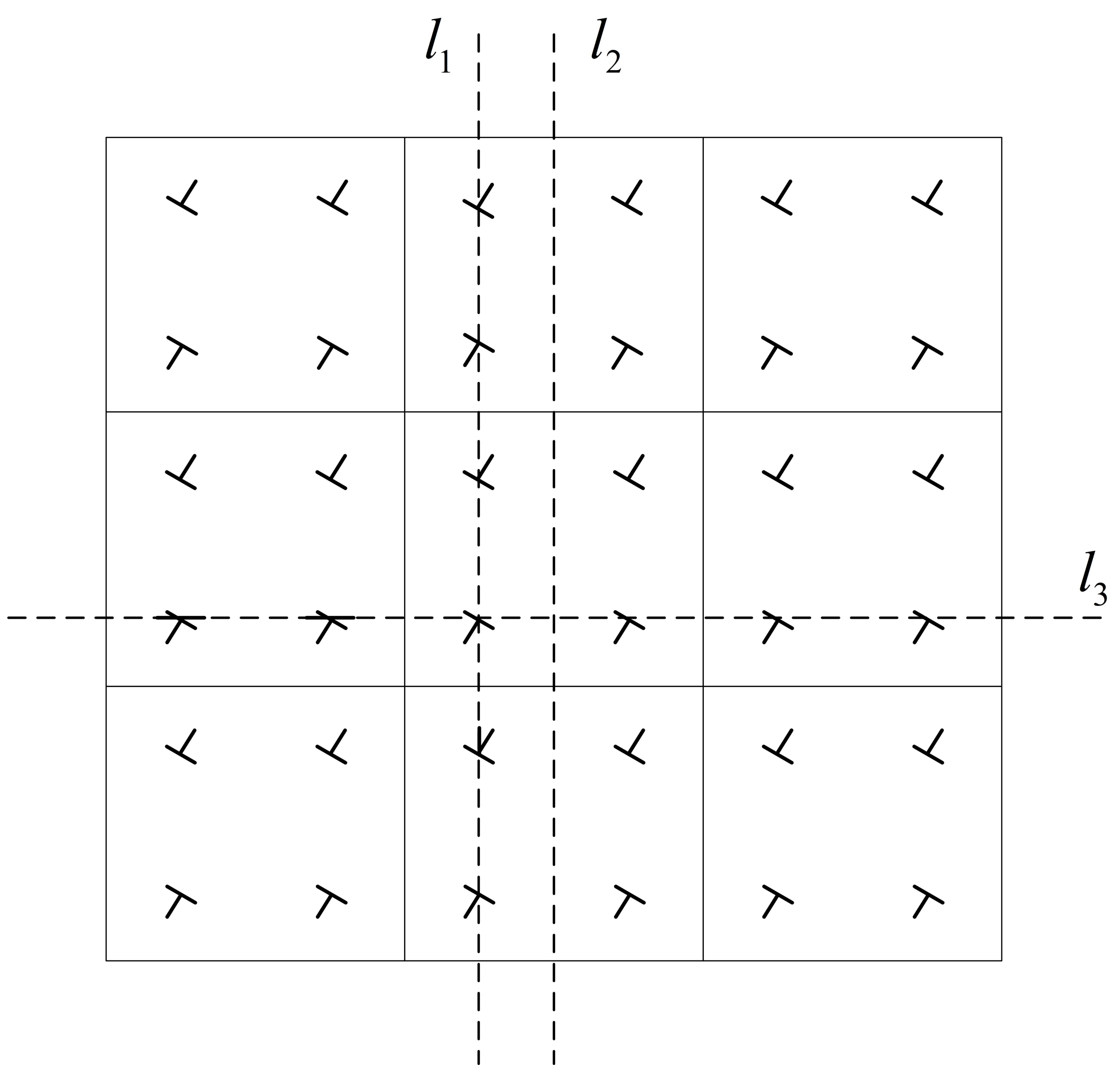}
	\caption{Illustration of periodically repeated domains, where each periodic domain contains a quadrupole edge dislocation.}
	\label{fig:direction}
\end{figure}

To verify the accuracy, we examine the strain field in three representative directions, as illustrated in Figure~\ref{fig:direction}. These directions are chosen based on the distribution of the strain field. Specifically, we consider the distribution of the strain field $\varepsilon_{xx}$ along the directions $l_1$ and $l_2$, and the distribution of $\varepsilon_{xy}$ along the direction $l_3$. We then compare the strain field in these directions with the theoretical results derived from continuum elasticity theory. Given the symmetry and periodicity of the lattice, it is essential to consider not only the four dislocation cores in the computational domain but also the contributions of dislocations from the surrounding lattice units to the strain field. The periodic extension is performed along the $x$ and $y$ directions, with calculations continuing until reaching the 100th layer. At this point, the error in the strain field between the last two layers is $1 \times 10^{-5}$.

\begin{figure}[htbp]
	\centering
	\subfloat[$\varepsilon_{xx}$ along $l_1$]{\includegraphics[width=.32\linewidth]{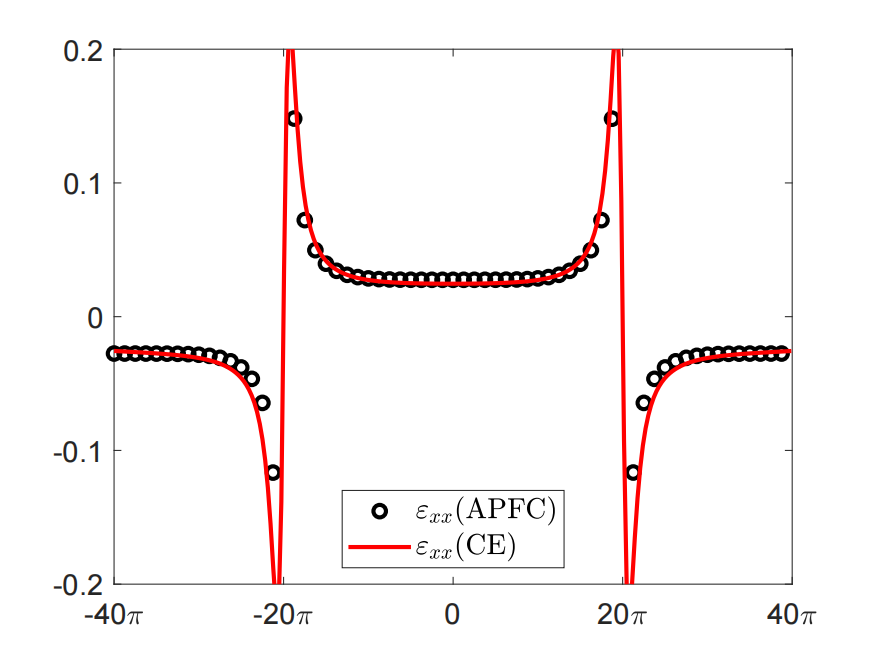}}~~
	\subfloat[$\varepsilon_{xx}$ along $l_2$]{\includegraphics[width=.32\linewidth]{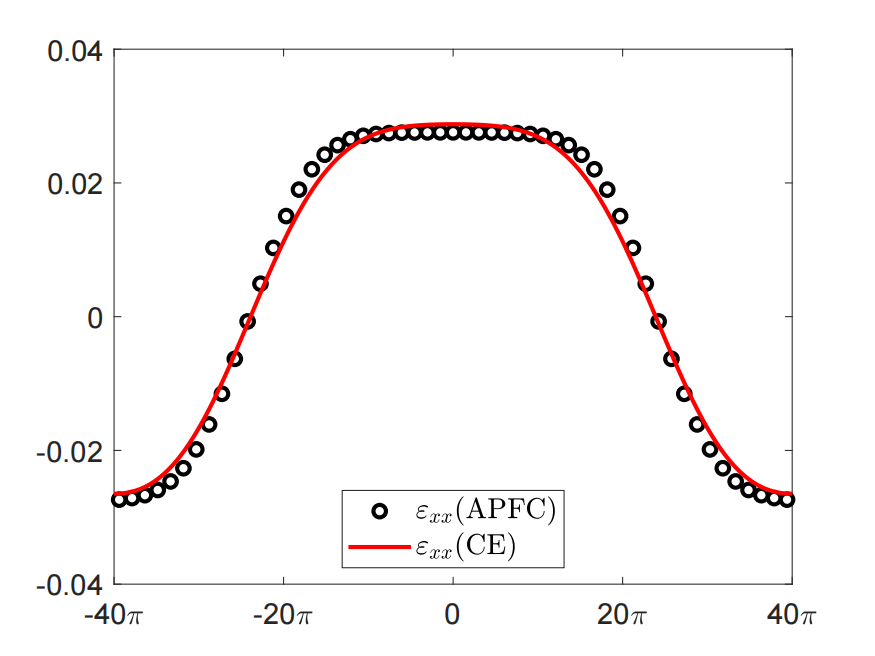}}~~
	\subfloat[$\varepsilon_{xy}$ along $l_3$]{\includegraphics[width=.32\linewidth]{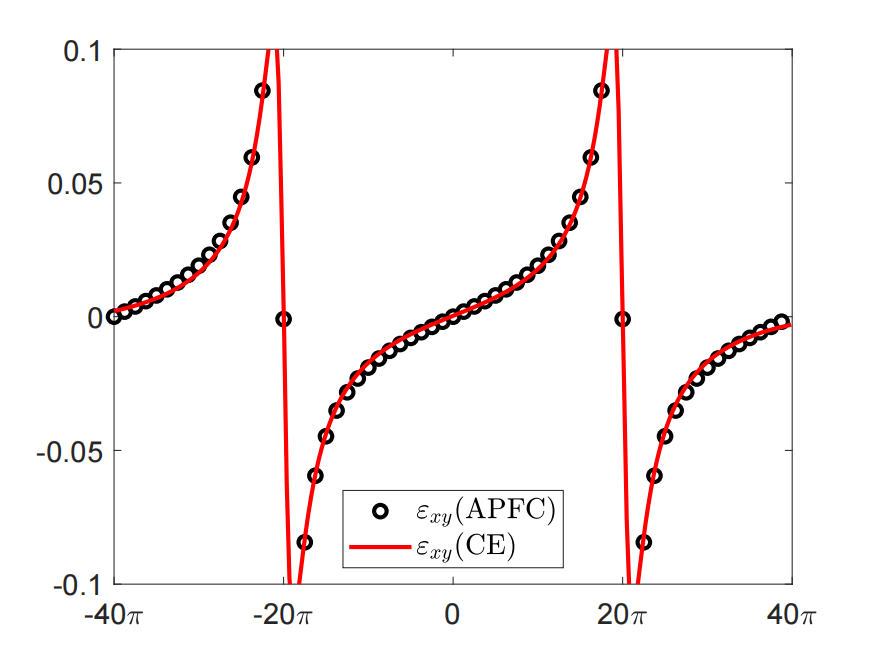}}
	\caption{Triangular lattice: Strain fields compared with CE along different lines. The red solid lines represent the results from CE theory, while the black circles represent the results from the APFC model~\eqref{eqn:apfcdynamics}.}
	\label{fig:APFC-CE2D}
\end{figure}

As shown in Figure~\ref{fig:APFC-CE2D}, the strain field from the APFC model~\eqref{eqn:apfcdynamics} is in good agreement with those obtained from continuum elasticity. Notably, we observe variations around the dislocation cores along the $l_1$ and $l_3$ directions, which reflects the expected strain field behavior in the presence of dislocations. This demonstrates the effectiveness of the Fourier pseudospectral method in solving the APFC model~\eqref{eqn:apfcdynamics}, yielding results that closely align with elasticity theory.

\begin{figure}[htbp]
	\centering
	\includegraphics[width=.5\linewidth]{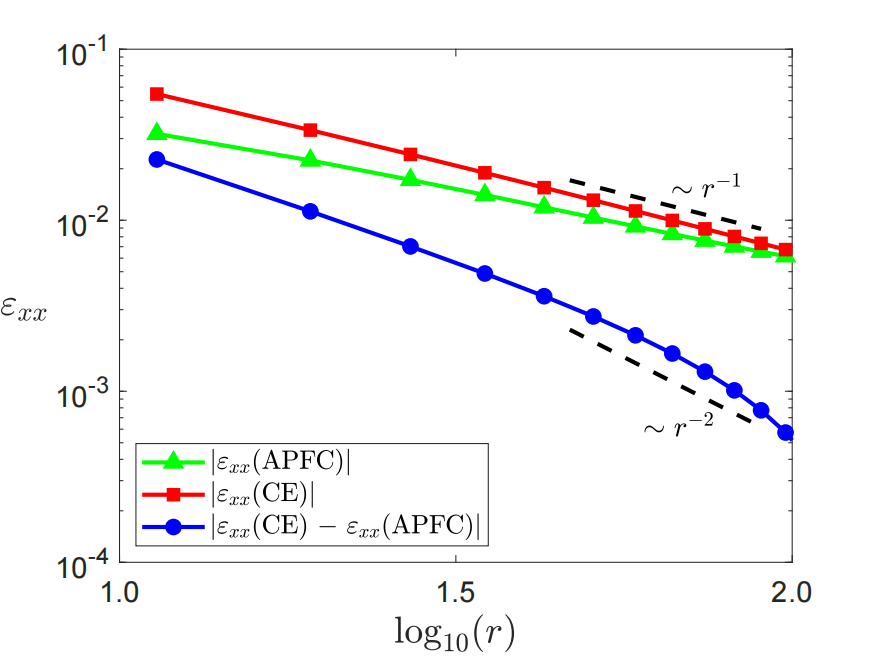}
	\caption{Triangular lattice: The decay of the strain field $\varepsilon_{xx}$ along the $l_1$ direction as a function of the distance $r$ from the dislocation core. The red line represents the results obtained from CE theory, while the green line shows the results from the APFC model~\eqref{eqn:apfcdynamics}. The blue line represents the error between them.}
	\label{Decay2D-xx}
\end{figure}

\paragraph{Far-field decay.} 
To verify the accuracy of the Fourier pseudospectral method for solving the APFC model~\eqref{eqn:apfcdynamics}, we consider a larger computational domain with a side length of $640\pi$. We focus on the case of $\varepsilon_{xx}$ in the direction $l_1$, noting that the strain field in other directions yields similar results. We demonstrate the results by comparing the decay of $\varepsilon_{xx}$ as a function of the distance $r$ from the dislocation core. As shown in Figure~\ref{Decay2D-xx}, both $\varepsilon_{xx}(\text{APFC})$ and $\varepsilon_{xx}(\text{CE})$, representing the results obtained using the APFC model~\eqref{eqn:apfcdynamics} and continuum elasticity (CE) theory, respectively, decay like $r^{-1}$. The error between them decays as $r^{-2}$, which is consistent with the results shown in~\cite{2019Closing}.

\subsection{Three-dimensional BCC lattice}
\label{sec:numericalexperiments:3d}

For the three-dimensional BCC lattice described in Example~\ref{ex} , we first let $\phi_j = \phi_0$ represent the amplitude of the equilibrium crystal. By minimizing the energy functional given in \eqref{eqn:energyfunctionalineta}, we obtain $\phi_0$ as follows:
\begin{equation*}
	\phi_0 = \frac{2\gamma + \sqrt{4\gamma^2 - 45vB_0}}{45v}.
\end{equation*}
Further computational details can be found in Appendix~\ref{sec:sub:phi0}.

A cubic computational domain $L_x \times L_y \times L_z$, with edge lengths $L_x = L_y = L_z = 60\pi$, is utilized for numerical simulations. To simulate crystal dislocation, shear stress is applied as in the two-dimensional case. Only horizontal shear stress parallel to the $xy$ plane is applied, with no force exerted in the $z$ direction. Based on $\eta_j = \phi_0e^{\imath\bm{k}_j\cdot \bm{u}(\bm{r})}$, the initial condition $\eta_{j,0}$ of the amplitude functions can be determined by defining $\bm{u}(\bm{r})$ as a displacement to the lattice. Given $\bm{r} := x\widehat{\bm{x}} + y\widehat{\bm{y}} + z\widehat{\bm{z}}$ representing the spatial position vector, the definition of $\bm{u}(\bm{r})$ is then given by
\begin{equation}
	\bm{u}(\bm{r})=\left\{
	\begin{aligned}
		-u_x \widehat{\bm{x}} -u_y \widehat{\bm{y}}&,& \frac{L_z}{4} < z < \frac{3L_z}{4} \\
		+u_x \widehat{\bm{x}} +u_y \widehat{\bm{y}} &,& {\rm elsewhere}
	\end{aligned}
	\right.,
\end{equation}
where $u_x = (a_x/L_x)x$, $u_y = (a_y/L_y)y$ and $u_z = 0$ with $a_x = a_y = 2\pi\sqrt{2}$. With this choice, a strain of 
\begin{equation}
	\varepsilon_x = \pm a_x/L_x, \quad \varepsilon_y = \pm a_y/L_y,
	\label{3Dstrain}
\end{equation}
is applied to the corresponding regions.

\begin{figure}[htbp]
	\centering
	\includegraphics[width=.5\linewidth]{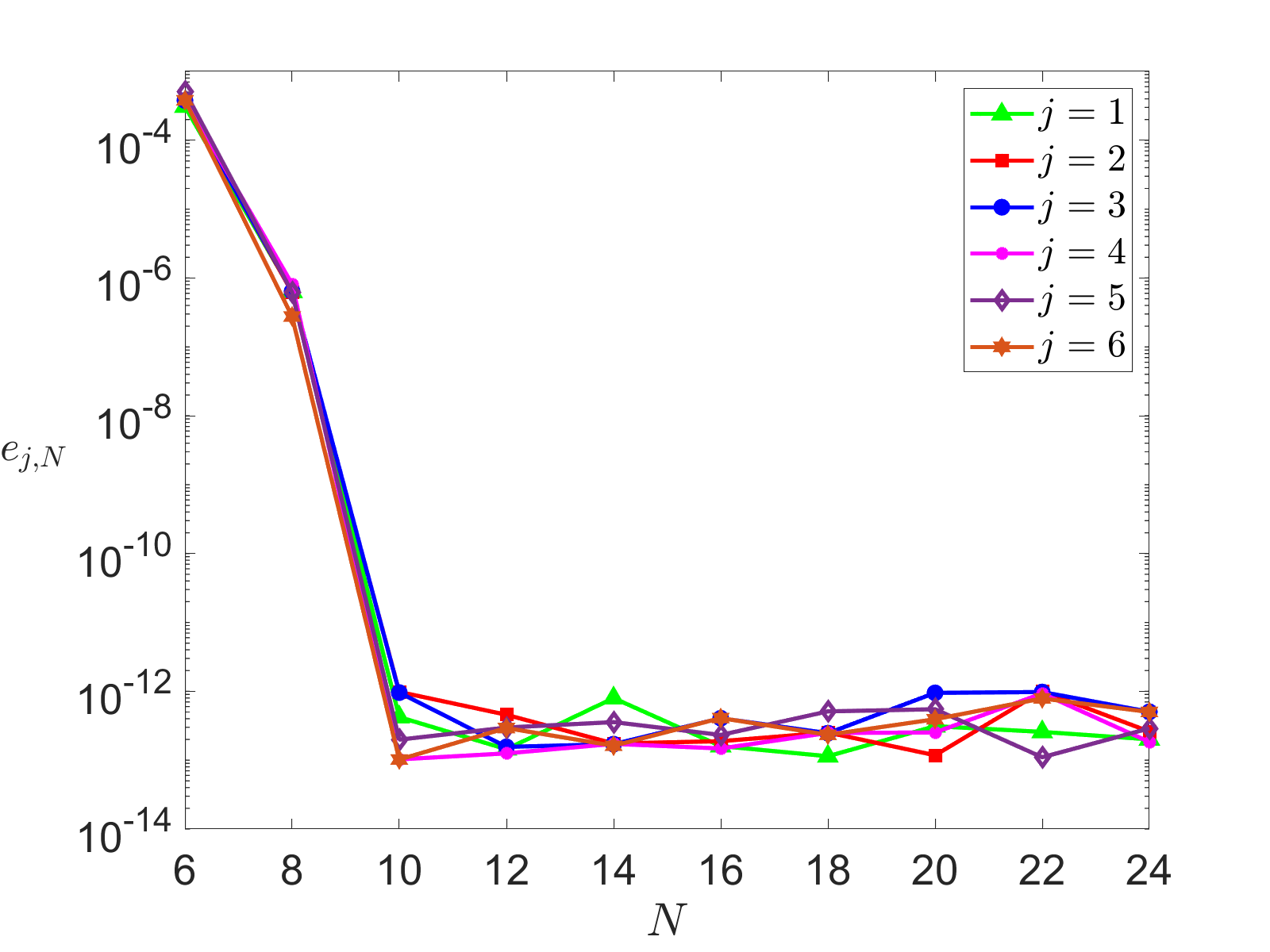}
	\caption{BCC lattice: numerical error $e_{j,N}$ for $j=1,\ldots,6$ as a function of $N$.}
	\label{fig:error-3D}
\end{figure}

Similar to the two-dimensional case, the numerical accuracy for the BCC lattice is evaluated using the same Fourier pseudospectral setting ($\tau = 1 \times 10^{-4}$, $h = \pi /128$, $T = 6$). The reference solution $\eta_j^{*}(\bm{r},T)$ and the numerical error $e_{j,N}$ are defined analogously to~\eqref{eq:exact} and~\eqref{eq:error} with $j=1,\ldots,6$ and $d=3$. As shown in Figure~\ref{fig:error-3D}, the errors $e_{j,N}$ for the BCC lattice display spectral convergence with respect to $N$, in agreement with the theoretical results in Section~\ref{sec:numerical methods}.

Table~\ref{tab:3Dtime} shows the computational time required to solve the model with varying $N$ in the three-dimensional case, using the same time step as in the two-dimensional scenario. Although the computation time has increased, the method retains high efficiency, scaling approximately as $O(N^{2.24})$. It is worth noting that our code has not been optimized; with proper optimization, the CPU computation time could be significantly reduced, leading to even better performance. This further underscores the feasibility of the Fourier pseudospectral method for efficient computations, as discussed in Section~\ref{sec: Implementation of Fourier pseudospectral method}.

\begin{table}[htbp]
	\caption{BCC lattice: CPU time (s) vs. $N$.}
	\label{tab:3Dtime}       
	\begin{tabular}{lllllll}
		\hline\noalign{\smallskip}
		$N$&32&64&128&256&512&1024 \\
		\noalign{\smallskip}\hline\noalign{\smallskip}
		CPU time (s) &	21.032	&	62.721&	319.545&	1854.294&	13801.318&	31014.621\\
		\noalign{\smallskip}\hline
	\end{tabular}
\end{table}

\subsubsection{Density $n$ and amplitude parameter $A^2$}
We employ the same parameters used in Section~\ref{sec:numericalexperiments:2d} to solve the APFC model~\eqref{eqn:apfcdynamics}. Based on how we apply the initial values, we study the distribution of the parameter $A^2$ defined in \eqref{eqn:A2} on the planes $z = L_z/4$ and $z = 3L_z/4$. The distributions on these planes are similar; thus, we present the case for the $z = L_z/4$ plane without loss of generality.

\begin{figure}[htbp]
	\centering
	\subfloat[$n$ of the perfect lattice]{\label{fig:A2-3D-a}
		\hspace{-2mm}
		\includegraphics[width=.23\linewidth]{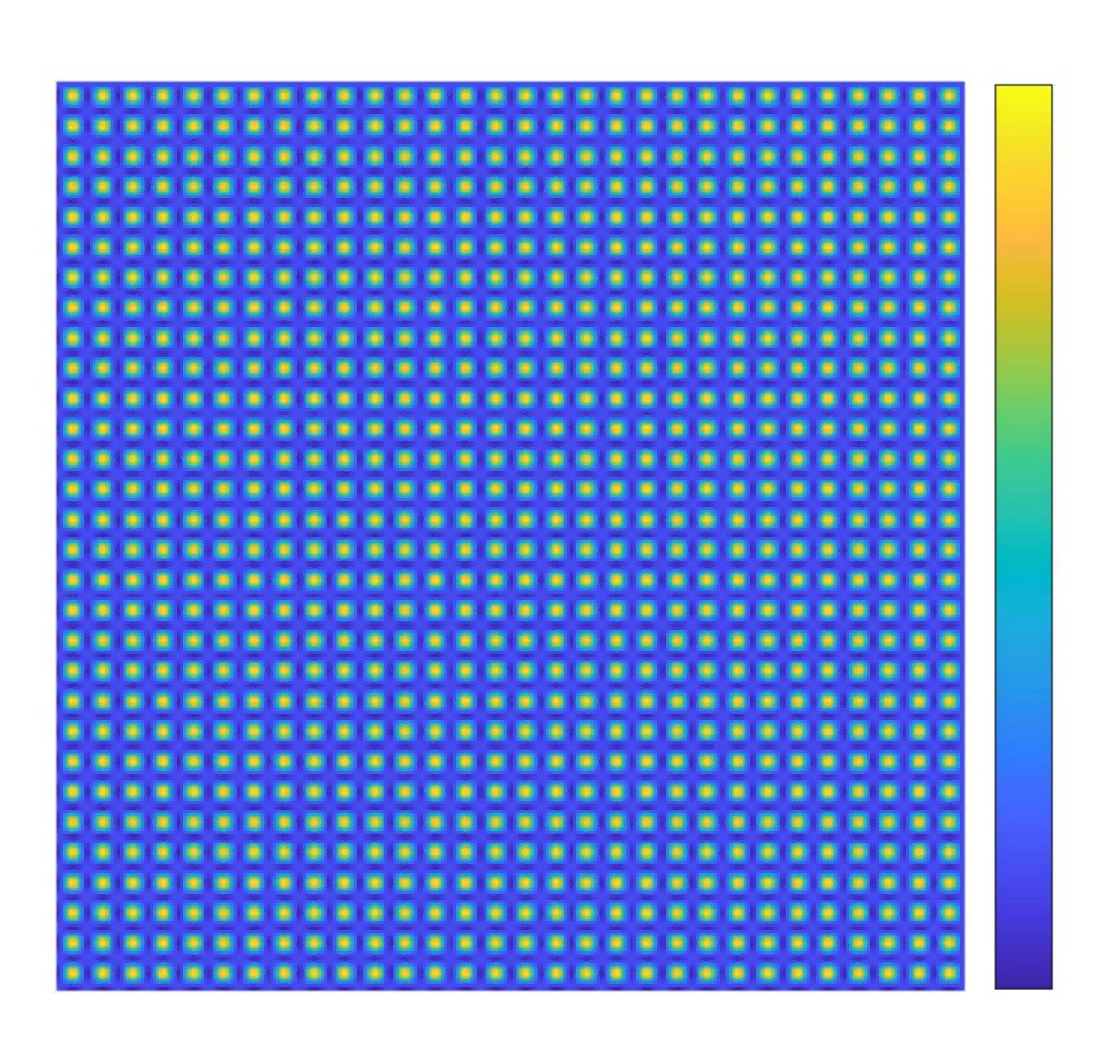}}~~
	\subfloat[$n$ of the defect lattice]{\label{fig:A2-3D-b}
		\hspace{-2mm}
		\includegraphics[width=.45\linewidth]{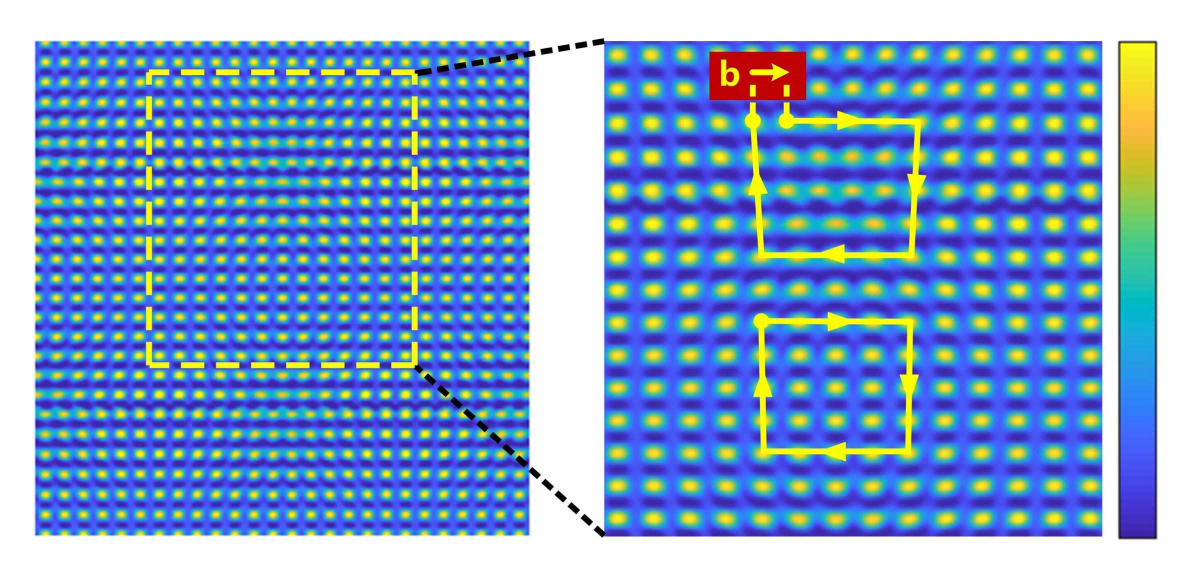}}~~
	\hspace{-2mm}
	\subfloat[$A^2$ on $z=L_z/4$ plane]{\label{fig:A2-3D-c}\includegraphics[width=.29\linewidth]{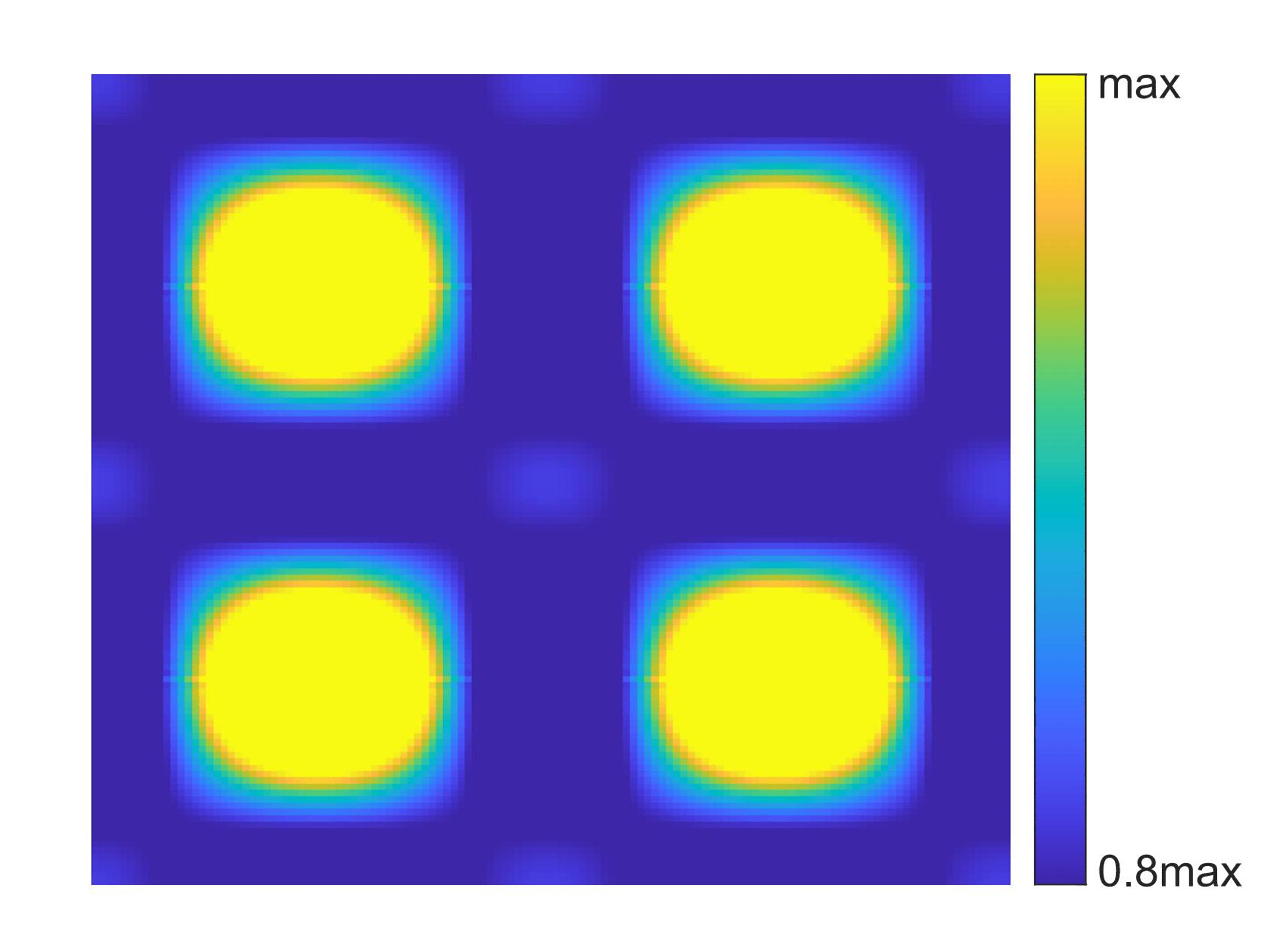}}
	\caption{BCC lattice: The illustration of the density $n$ defined in \eqref{eqn:apfcdensity} for the perfect lattice and defect lattice, and the parameter $A^2$ on the $z = L_z/4$ plane. Both are colored by their respective values.}
	\label{fig:A2-3D}
\end{figure}

Figure~\ref{fig:A2-3D} visualizes the density $n$ on the $yz$-plane with $x = L_x/4$ and the parameter $A^2$ on the $z = L_z/4$ plane, illustrating the distribution of dislocations as the system stabilizes. Figure~\ref{fig:A2-3D-a} and Figure~\ref{fig:A2-3D-b} display the density $n$ of the perfect lattice and the defect lattice, which can be reconstructed from \eqref{eqn:apfcdensity}. Similar to the two-dimensional situation, the area near the defect is extracted in Figure~\ref{fig:A2-3D-b}, and the Burgers vector can be clearly seen. We focus on the regions where $A^2$ decreases in Figure~\ref{fig:A2-3D-c}. The navy blue area corresponds to the region where $A^2 < 0.8 \max(A^2)$, indicating the presence of defects as $A^2$ significantly decreases. The distribution in the figure demonstrates the property of the parameter $A^2$: it remains constant within ordered phases but decreases in the presence of defects.

\subsubsection{Strain field and comparison with continuum elasticity}
\label{sec:sub:strain_field_3D}

We employ the least squares method to solve the system \eqref{N-eqs}. After some straightforward calculations, the least squares solution of \eqref{N-eqs} in the context of the Moore-Penrose inverse is expressed as follows:
\begin{equation}
	\bm{u}(\bm{r})=\left\{
	\begin{array}{lr}
		u_x = \frac{\sqrt{2}}{4}(\varphi_1 + \varphi_2 + \varphi_5 - \varphi_6)\\
		u_y =  \frac{\sqrt{2}}{4}(\varphi_1 + \varphi_3 + \varphi_4 - \varphi_5)\\
		u_z =  \frac{\sqrt{2}}{4}(\varphi_2 + \varphi_3 - \varphi_4 + \varphi_6)
	\end{array}
	\right..
	\label{u3D}
\end{equation}

\paragraph{Strain field.} 

According to the shear stress applied in \eqref{3Dstrain} and the defect distribution shown in Figure~\ref{fig:A2-3D}, we choose to study the strain field distribution on the $xz$-plane and $yz$-plane. Given the symmetric structure of the BCC lattice, their distributions should be the same. Without loss of generality, we consider the strain field on the $yz$-plane with $x = L_x/4$. Similar to the two-dimensional case, the strain field $\varepsilon_{yz}$ can be computed using the following formulation:
\begin{equation*}
	\varepsilon_{yz} = \frac{\sqrt{2}}{8}\left[\left(\frac{\partial \varphi_1}{\partial y} + \frac{\partial \varphi_3}{\partial y} + \frac{\partial \varphi_4}{\partial y} - \frac{\partial \varphi_5}{\partial y}\right) + \left(\frac{\partial \varphi_2}{\partial z} + \frac{\partial \varphi_3}{\partial z} - \frac{\partial \varphi_4}{\partial z} + \frac{\partial \varphi_6}{\partial z}\right)\right].
\end{equation*}

\begin{figure}[htbp]
	\centering
	\includegraphics[width=.5\linewidth]{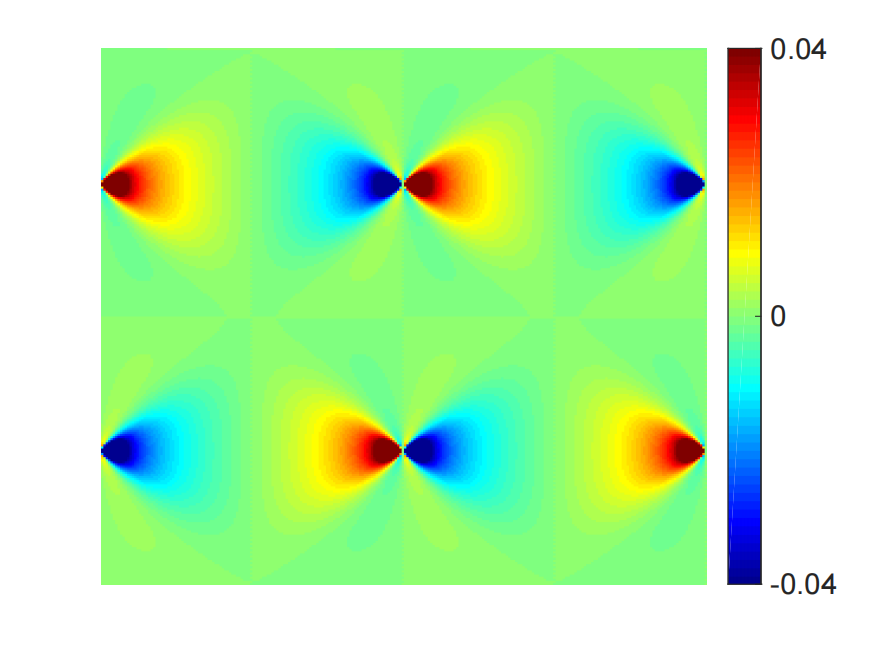}
	\caption{BCC lattice: The illustration of $\varepsilon_{yz}$.}
	\label{fig:strainfieldyz}
\end{figure}

Figure~\ref{fig:strainfieldyz} shows the strain field $\varepsilon_{yz}$ for the three-dimensional BCC lattice. On this plane, we observe parallel dislocations, and the distribution of the field at each defect core aligns with the theoretical expectations discussed in prior work~\cite{anderson2017theory}. 

\paragraph{Comparison with continuum elasticity theory.} 

To verify the accuracy of the APFC solution, we examine the strain field $\varepsilon_{yz}$ along the $z = L_z/4$ direction on the $yz$-plane and compare the strain field from the APFC model~\eqref{eqn:apfcdynamics} with the theoretical results derived from continuum elasticity theory.

In the three-dimensional case, based on the distribution of defects shown in Figure~\ref{fig:A2-3D}, we need to consider two sets of dislocations, with both dislocation cores located at $\bm{x}_c = (0,0,0)$ and the Burgers vector perpendicular to each other, aligned parallel to the $\bm{x}$-axis and $\bm{y}$-axis, respectively. Similar to the two-dimensional case, the strain field $\varepsilon_{yz}$ can be obtained as follows:
\begin{equation}
	\begin{split}
		\varepsilon_{yz} &= \frac{1+v}{E}\sigma_{yz}.
	\end{split}
\end{equation}

\begin{figure}[htbp]
	\centering
	\includegraphics[width=.5\linewidth]{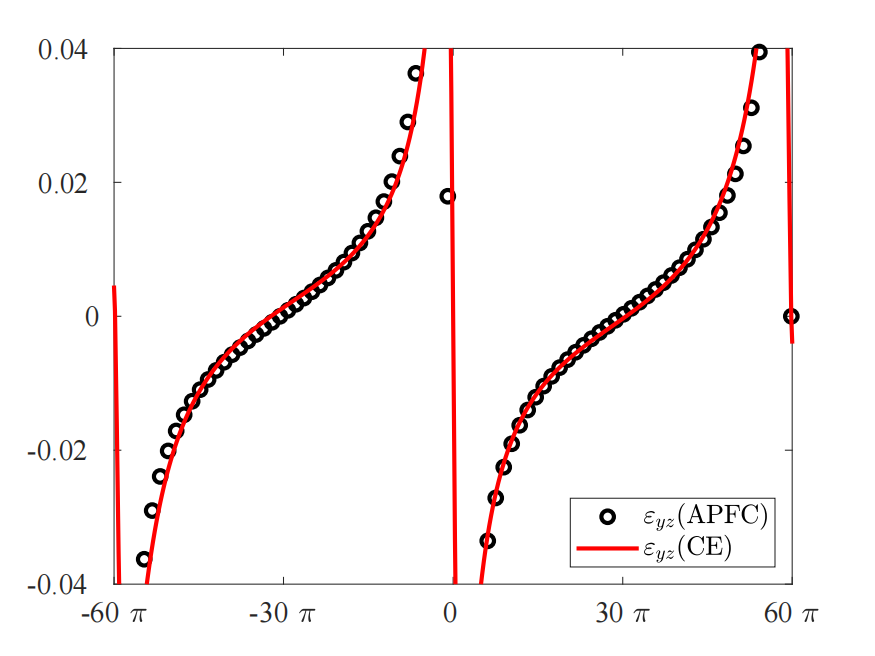}
	\caption{BCC lattice: Strain field $\varepsilon_{yz}$ compared with CE along the $z = L_z/4$ direction on the $yz$-plane. The red solid lines represent the results from CE theory, while the black circles represent the results from the APFC model~\eqref{eqn:apfcdynamics}.}
	\label{fig:APFC-CE3D}
\end{figure}

As shown in Figure~\ref{fig:APFC-CE3D}, the strain field obtained from the APFC model~\eqref{eqn:apfcdynamics} closely matches those derived from continuum elasticity (CE). Notably, there is significant numerical variation around the dislocation cores, which is consistent with the expected strain field behavior in these regions. This outcome reaffirms the effectiveness of the Fourier pseudospectral method in solving the three-dimensional APFC model~\eqref{eqn:apfcdynamics}, producing results that align well with continuum elasticity theory.

\paragraph{Far-field decay.} 

\begin{figure}[htbp]
	\centering
	\includegraphics[width=.5\linewidth]{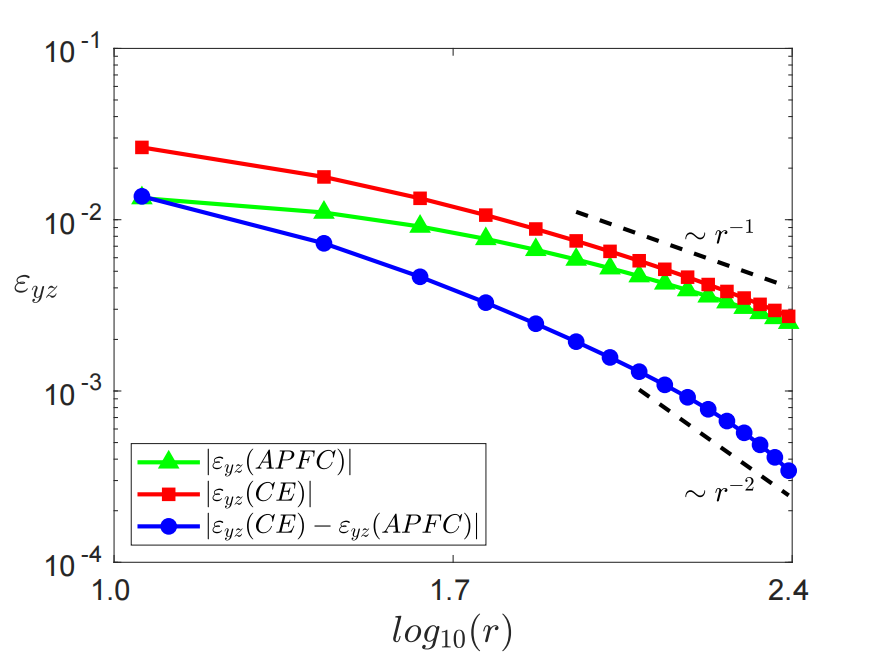}
	\centering
	\caption{BCC lattice: The decay of the strain field $\varepsilon_{yz}$  along $z=L_z/4$ direction on the $yz$-plane in terms of the distance $r$ from the dislocation core. The red line represents the results from the CE theory while the green line shows the results by using the APFC model~\eqref{eqn:apfcdynamics}. Moreover, the blue line represents the error between them.}
	\label{fig:Decay3D}
\end{figure}

Similar to the two-dimensional case, we consider a larger computational domain with a side length of 640$\pi$ to verify the accuracy and computational efficiency of the numerical method for solving the APFC model~\eqref{eqn:apfcdynamics}, as well as to verify the decay of the strain field away from the dislocation core. We demonstrate the results by comparing the decay of $\varepsilon_{yz}$ in terms of the distance $r$ from the dislocation core. As illustrated in Figure~\ref{fig:Decay3D}, both the strain field $\varepsilon_{yz}(\rm{APFC})$ and $\varepsilon_{yz}(\rm{CE})$, corresponding to the APFC model~\eqref{eqn:apfcdynamics} and CE theory, exhibit a decay behavior proportional to $r^{-1}$. Moreover, the error between them decays with a rate of $r^{-2}$. This behavior mirrors the trends observed in the two-dimensional case, underscoring the robust accuracy of our numerical method in effectively solving the APFC model~\eqref{eqn:apfcdynamics}.

\section{Conclusion}
\label{sec:conclusion}

In this paper, we have demonstrated the Fourier pseudospectral method for solving the APFC model in dislocation simulations. By leveraging the system's inherent periodicity and applying periodic boundary conditions, the Fourier pseudospectral method offers exceptional accuracy and computational efficiency. Our numerical experiments on two-dimensional triangular lattices and three-dimensional body-centered cubic lattices successfully optimized edge dislocation geometries, yielding strain fields consistent with the predictions of continuum elasticity theory. These results highlight the potential of this method in advancing the simulation of complex crystalline defects. Our findings suggest broad applicability of this numerical approach to other crystalline defects. Several open problems warrant further investigation:
\begin{itemize}
	\item Grain-boundary dynamics: Extending the APFC model to grain boundary problems by incorporating a rotation matrix within amplitude functions could enable the simulation of grain boundary evolution. This extension, crucial for studying deformation and rotation in polycrystalline systems, will be a focus of our future work.
	\item Closing the gap between the APFC and atomistic models: Aligning the elastic parameters of the APFC and atomistic models is key to establishing a connection between them. Future research should explore coupling methods similar to classical atomistic-to-continuum techniques to effectively bridge these models~\cite{2019Closing}. 
	\item Uncertainty quantification of the APFC model: Uncertainty quantification (UQ) is essential in computational materials science to enhance model reliability and guide experimental design. For the APFC model, understanding how parameter uncertainties impact predictions of material properties will be critical, especially when complex structures emerge from varying initial conditions.
	\item Machine learning in the APFC model: The APFC model encompasses parameters that necessitate calibration for accurately reproducing experimental observations. One promising approach is to utilize machine learning techniques to develop a precise and robust model for the phase field crystal model, facilitating the prediction of crystal structures and enhancing our comprehension of material defect states.
\end{itemize}

\appendix
\section{Introduction to Bravais Lattices and Reciprocal Lattices}
\label{sec:Reciprocal Lattice}
In this section, we provide an introduction to Bravais lattices and reciprocal lattices, as well as their mathematical formulations. For any $d$-dimensional periodic structure $F(\bm{r})$, where $\bm{r} \in \mathbb{R}^d$, the repeated structural unit is called a unit cell. 

The Bravais lattice vector is defined as
\begin{equation*}
	\bm{R}_{\bm{n}} = n_1 \bm{e}_1 + n_2 \bm{e}_2 + \ldots + n_d \bm{e}_d, \quad n_i \in \mathbb{Z},
\end{equation*}
where $\bm{n} = (n_1, n_2, \ldots, n_d)$ is a $d$-dimensional vector. Given the primitive vectors $(\bm{e}_1, \bm{e}_2, \ldots, \bm{e}_d)$, the reciprocal vectors $(\bm{e}_1^{*}, \bm{e}_2^{*}, \ldots, \bm{e}_d^{*})$ satisfy the equation:
\begin{equation*}
	\bm{e}_i \cdot \bm{e}_j^{*} = 2\pi \delta_{ij}.
\end{equation*}
The reciprocal lattice vector is given by:
\begin{equation*}
	\bm{G}_{\bm{m}} = m_1 \bm{e}_1^{*} + m_2 \bm{e}_2^{*} + \ldots + m_d \bm{e}_d^{*}, \quad m_i \in \mathbb{Z},
\end{equation*}
where $\bm{m} = (m_1, m_2, \ldots, m_d)$ is a $d$-dimensional integer vector.

One of the most important properties of the reciprocal lattice vector is that the plane wave $\{ e^{\imath \bm{G}_{\bm{m}} \cdot \bm{r}} \}$ can form a set of basis functions for any periodic function $f \in L^2(\mathbb{T}_N^d)$. This can be expressed as
\begin{equation}
	f(\bm{r}) = \sum_{\bm{m}} \widehat{f}_{\bm{m}} e^{\imath {\bm{G}_{\bm{m}}} \cdot {\bm{r}}},
	\label{eqn:wave expansion}
\end{equation}
where $\widehat{f}_{\bm{m}}$ are Fourier coefficients. For any $\bm{R}_{\bm{n}}$ in the Bravais lattice, the structure is invariant under lattice translation:
\begin{equation}
	f(\bm{r} + \bm{R}_{\bm{n}}) = f(\bm{r}).
	\label{eqn:invariant under lattice translation}
\end{equation}
Expanding the functions on both sides of~\eqref{eqn:invariant under lattice translation} as in~\eqref{eqn:wave expansion}, we have
\begin{equation*}
	\sum_{\bm{m}} \widehat{f}_{\bm{m}} e^{\imath {\bm{G}_{\bm{m}}} \cdot {\bm{r}}} = \sum_{\bm{m}} \widehat{f}_{\bm{m}} e^{\imath {\bm{G}_{\bm{m}}} \cdot (\bm{r} + \bm{R}_{\bm{n}})} = \sum_{\bm{m}} \widehat{f}_{\bm{m}} e^{\imath {\bm{G}_{\bm{m}}} \cdot {\bm{r}}} e^{\imath {\bm{G}_{\bm{m}}} \cdot {\bm{R}_{\bm{n}}}}.
\end{equation*}
Note that $e^{\imath {\bm{G}_{\bm{m}}} \cdot {\bm{R}_{\bm{n}}}} = 1$ holds only if the following equation is satisfied:
\begin{equation}
	\bm{G}_{\bm{m}} \cdot \bm{R}_{\bm{n}} = 2\pi N, \quad N \in \mathbb{Z}.
\end{equation}
Hence, equation~\eqref{eqn:wave expansion} can be rewritten as:
\begin{equation}
	\begin{aligned}
		f(\bm{r}) &= \sum_{\bm{m}} \widehat{f}_{\bm{m}} e^{\imath {\bm{G}_{\bm{m}}} \cdot {\bm{r}}} \\
		&= \sum_{\bm{k} \in K_N^d} \widehat{f}_{\bm{k}} e^{\imath \left( \frac{2\pi}{\bm{R}_{\bm{n}}} \cdot \bm{k} \right) \cdot {\bm{r}}} \\
		&= \sum_{\bm{k} \in K_N^d} \widehat{f}_{\bm{k}} e^{\imath {(B \bm{k}) \cdot {\bm{r}}}},
		\label{eqn:periodic expansion}
	\end{aligned}
\end{equation}
where $B = 2\pi/\bm{R}_{\bm{n}}$ denotes the transformation matrix from physical space to reciprocal space.

\section{Auxiliary Results}
\label{sec:appendix}

\subsection{The derivation of \eqref{eqn:apfcdynamics-rightterm}}
\label{sec:sub:apd}

We first derive \eqref{eqn:apfcdynamics-rightterm}. We begin by computing $A^2$ and $A^4$:
\begin{align*}
	A^2 &= 2\sum_{j=1}^M\lvert\eta_j\rvert^2 = 2\sum_{j=1}^M\eta_j\eta_j^*, \\
	A^4 &= 4\left(\sum_{j=1}^M\lvert\eta_j\rvert^2\right)^2 = 4\left(\sum_{j=1}^M\eta_j\eta_j^*\right)^2.
\end{align*}

Substituting these into \eqref{eqn:energyfunctionalineta}, we obtain
\begin{equation}
	\begin{aligned}
		F &= \frac{1}{\lvert\Omega\rvert}\int_{\Omega} \Big[\frac{B_0}{2}A^2 + \frac{3v}{4}A^4 + \sum_{j=1}^M \left(B_x\lvert\varsigma_j \eta_j\rvert^2 - \frac{3v}{2} \lvert\eta_j\rvert^4\right) + f^s({\eta_j},{\eta_j^*})\Big] \,{\rm d}\bm{r} \\
		&= \frac{1}{\lvert\Omega\rvert}\int_{\Omega} \Big[B_0\sum_{j=1}^M\lvert\eta_j\rvert^2 + 3v\left(\sum_{j=1}^M\eta_j\eta_j^*\right)^2 + \sum_{j=1}^M \left(B_x\lvert\varsigma_j \eta_j\rvert^2 - \frac{3v}{2} \lvert\eta_j\rvert^4\right)\Big] \,{\rm d}\bm{r}\\
		&\quad+ \frac{1}{\lvert\Omega\rvert}\int_{\Omega} f^s({\eta_j},{\eta_j^*})\,{\rm d}\bm{r} .
	\end{aligned}
	\label{eqn:Fexpan}
\end{equation}

The partial functional derivative $\delta F/\delta \eta_j^*$ of $F$ with respect to $\eta_j$ can be calculated as follows. For any test function $\phi_j \in U$, where $U$ is a Banach space of functions $\eta_j$ on the computational domain $\Omega \in \mathbb{R}^d$, we have
\begin{align*}
	\left(\frac{\delta F}{\delta \eta_j^*}, \phi_j\right) &= \frac{d}{dt} F(\eta_1, \ldots, \eta_M, \eta_1^*, \ldots, \eta_j^* + t\phi_j, \ldots, \eta_M^*)\bigg\lvert _{t=0} \\
	&= \lim_{t \to 0} \frac{F(\eta_1, \ldots, \eta_M, \eta_1^*, \ldots, \eta_j^* + t\phi_j, \ldots, \eta_M^*)}{t} \\
	&\quad -\lim_{t \to 0} \frac{F(\eta_1, \ldots, \eta_M, \eta_1^*, \ldots, \eta_M^*)}{t} \\
	&= \lim_{t \to 0} \frac{1}{t\lvert\Omega\rvert }\int_{\Omega} \left[B_0 t \eta_j \phi_j + 3v\left((A^2 - \eta_j \eta_j^*) + \eta_j t \phi_j\right)(t \eta_j \phi_j) \right] \,{\rm d}\bm{r}\\
	&\quad + \lim_{t \to 0} \frac{1}{t\lvert\Omega\rvert }\int_{\Omega} B_x t \varsigma_j^2 \eta_j \phi_j - \frac{3v}{2} t \eta_j^2 \phi_j^2 \,{\rm d}\bm{r} + \left(\frac{\delta f^s}{\delta \eta_j^*}, \phi_j\right) \\
	&= \frac{1}{\lvert\Omega\rvert }\int_{\Omega} \left[B_0 \eta_j \phi_j + 3v\left(A^2 - \lvert\eta_j\rvert ^2\right) \eta_j \phi_j + B_x \varsigma_j^2 \eta_j \phi_j\right] \,{\rm d}\bm{r} + \left(\frac{\delta f^s}{\delta \eta_j^*}, \phi_j\right) \\
	&= \left(\left[B_0 + 3v\left(A^2 - \lvert\eta_j\rvert^2\right) + B_x \varsigma_j^2\right] \eta_j, \phi_j\right) + \left(\frac{\delta f^s}{\delta \eta_j^*}, \phi_j\right).
\end{align*}

According to the fundamental lemma of the calculus of variations, for any $\phi_j$,
\begin{align*}
	\left(\frac{\delta F}{\delta \eta_j^*}, \phi_j\right) &= \left(\left[B_0 + B_x(\nabla^2 + 2\imath \bm{k}_j \cdot \nabla)^2 + 3v(A^2 - \lvert\eta_j\rvert^2)\right] \eta_j, \phi_j\right) + \left(\frac{\delta f^s}{\delta \eta_j^*}, \phi_j\right).
\end{align*}

Thus, we obtain
\begin{align}
	\frac{\delta F}{\delta \eta_j^*} &= \left[B_0 + B_x(\nabla^2 + 2\imath \bm{k}_j \cdot \nabla)^2 + 3v(A^2 - \lvert\eta_j\rvert^2)\right] \eta_j + \frac{\delta f^s}{\delta \eta_j^*}.
	\label{eqn:proofrightterm}
\end{align}

\subsubsection{The reciprocal-space vectors and ${\delta f^s}/{\delta \eta_j^*}$ defined in the Example~\ref{ex}}

In Example~\ref{ex}, the reciprocal-space vectors used to characterize the crystal symmetry and $\frac{\delta f^s}{\delta \eta_j^*}$ are provided according to the selected lattice symmetry as follows:

\begin{enumerate}
	\item For two-dimensional triangular lattices with $M=3$, the reciprocal-space vectors are
	\begin{equation*}
		\bm{k}_1 = \left(-\frac{\sqrt{3}}{2}, -\frac{1}{2}\right), \quad \bm{k}_2 = (0, 1), \quad \bm{k}_3 = \left(\frac{\sqrt{3}}{2}, -\frac{1}{2}\right).  
	\end{equation*}
	The complex polynomial functional in \eqref{eqn:apfcdynamics-rightterm} is given by
	\begin{align*}
		f^{s}_{\rm{tri}} = -2\gamma\left(\eta_1\eta_2\eta_3 + {\rm c.c.}\right).
	\end{align*}
	Following the derivation process of $\delta F/\delta \eta_j^*$, we take $j = 1$ as an example:
	
	\begin{align*}
		\left(\frac{\delta f^{s}_{\rm tri}}{\delta \eta_1^*}, \phi_1\right) &= \frac{d}{dt} f^{\rm tri}(\eta_1, \eta_2, \eta_3, \eta_1^* + t\phi_1, \eta_2^*, \eta_3^*)\bigg\lvert_{t=0} \\
		&= \lim_{t \to 0} \frac{f^{\rm tri}(\eta_1, \eta_2, \eta_3, \eta_1^* + t\phi_1, \eta_2^*, \eta_3^*)}{t}- \lim_{t \to 0} \frac{f^{\rm tri}(\eta_1, \eta_2, \eta_3, \eta_1^*, \eta_2^*, \eta_3^*)}{t} \\
		&= \lim_{t \to 0} -\frac{2\gamma}{t} \frac{1}{\lvert\Omega\rvert} \int_{\Omega} \left( (\eta_1^* + t\phi_1)\eta_2^*\eta_3^* - \eta_1^*\eta_2^*\eta_3^*\right) \,{\rm d}\bm{r} \\
		&= \lim_{t \to 0} -\frac{2\gamma}{t} \frac{1}{\lvert\Omega\rvert} \int_{\Omega} \eta_2^*\eta_3^* t\phi_1 \,{\rm d}\bm{r} \\
		&= \left(-2\gamma \eta_2^* \eta_3^*, \phi_1\right).
	\end{align*}
	Thus, we find
	\begin{equation*}
		\frac{\delta f^{\rm tri}}{\delta \eta_1^*} = -2\gamma \eta_2^* \eta_3^*.
	\end{equation*}
	In general,
	\begin{equation}
		\frac{\delta f^{s}_{\rm tri}}{\delta \eta_j^*} = -2\gamma \prod_{i \neq j}^3 \eta_i^*, \qquad j = 1, 2, 3.
	\end{equation}
	
	\item For the body-centered cubic (BCC) lattice with $M=6$, the reciprocal-space vectors are
	\begin{equation*}
		\begin{aligned}
			&\bm{k}_1 = (1, 1, 0),  &\bm{k}_2 &= (1, 0, 1),  &\bm{k}_3 &= (0, 1, 1), \\
			&\bm{k}_4 = (0, 1, -1), &\bm{k}_5 &= (1, -1, 0),  &\bm{k}_6 &= (-1, 0, 1).
		\end{aligned}
	\end{equation*}
	The complex polynomial functional in \eqref{eqn:apfcdynamics-rightterm} is given by
	\begin{equation*}
		\begin{aligned}
			f^{s}_{\rm bcc} = -&2\gamma\left(\eta_1^*\eta_2\eta_3 + \eta_2^*\eta_3\eta_5 + \eta_3^*\eta_1\eta_6 + \eta_4^*\eta_5^*\eta_6^* + {\rm c.c.}\right) \\ 
			&+ 6v\left(\eta_1\eta_3^*\eta_4^*\eta_5^* + \eta_2\eta_1^*\eta_5^*\eta_6^* + \eta_3\eta_2^*\eta_6^*\eta_4^* + {\rm c.c.}\right).
		\end{aligned}
	\end{equation*}
	
	Following a similar derivation process for $\delta F/\delta \eta_j^*$, we take $\eta_1^*$ as an example:
	\begin{align*}
		\left(\frac{\delta f^{\rm bcc}}{\delta \eta_1^*}, \phi_1\right) &= \frac{d}{dt} f^{\rm bcc}(\eta_1, \eta_2, \ldots, \eta_6, \eta_1^* + t\phi_1, \eta_2^*, \ldots, \eta_6^*)\bigg\lvert_{t=0} \\
		&= \lim_{t \to 0} \frac{f^{\rm bcc}(\eta_1, \eta_2, \ldots, \eta_6, \eta_1^* + t\phi_1, \eta_2^*, \ldots, \eta_6^*)}{t} \\
		&\quad-\lim_{t \to 0} \frac{f^{\rm bcc}(\eta_1, \eta_2, \ldots, \eta_6, \eta_1^*, \eta_2^*, \ldots, \eta_6^*)}{t} \\
		&= \lim_{t \to 0} -\frac{2\gamma}{t} \frac{1}{\lvert\Omega\rvert} \int_{\Omega} \left(\eta_2\eta_4 t\phi_1 + \eta_3\eta_6^* t\phi_1\right) \,{\rm d}\bm{r} \\
		& \quad + \lim_{t \to 0} \frac{6v}{t} \frac{1}{\lvert\Omega\rvert} \int_{\Omega} \left(\eta_2\eta_5^*\eta_6^* t\phi_1 + \eta_3\eta_4\eta_5 t\phi_1\right) \,{\rm d}\bm{r} \\
		&= \frac{1}{\lvert\Omega\rvert} \int_{\Omega} \left(-2\gamma(\eta_2\eta_4 + \eta_3\eta_6^*) + 6v(\eta_2\eta_5^*\eta_6^* + \eta_3\eta_4\eta_5)\right) \phi_1 \,{\rm d}\bm{r}.
	\end{align*}
	Thus, we find
	\begin{equation*}
		\frac{\delta f^{\rm bcc}}{\delta \eta_1^*} = -2\gamma(\eta_2\eta_4 + \eta_3\eta_6^*) + 6v(\eta_2\eta_5^*\eta_6^* + \eta_3\eta_4\eta_5).
	\end{equation*}
	
	In general,
	\begin{equation}
		\frac{\delta f^{\rm bcc}}{\delta \eta_i^*} = -2\gamma(\eta_k\eta_n^* + \eta_j\eta_l) + 6v(\eta_k\eta_l\eta_m + \eta_j\eta_m^*\eta_n^*).
		\label{eqn:fs123}
	\end{equation}
	
	For $\eta_4^*, \eta_5^*, \eta_6^*$, taking $\eta_4^*$ as an example:
	\begin{align*}
		\left(\frac{\delta f^{\rm bcc}}{\delta \eta_4^*}, \phi_4\right) &= \frac{d}{dt} f^{\rm bcc}(\eta_1, \eta_2, \ldots, \eta_6, \eta_1^*, \ldots, \eta_4^* + t\phi_4, \ldots, \eta_6^*)\bigg\lvert_{t=0} \\
		&= \lim_{t \to 0} \frac{f^{\rm bcc}(\eta_1, \eta_2, \ldots, \eta_6, \eta_1^*, \ldots, \eta_4^* + t\phi_4, \ldots, \eta_6^*)}{t} \\
		&\quad- \lim_{t \to 0} \frac{f^{\rm bcc}(\eta_1, \eta_2, \ldots, \eta_6, \eta_1^*, \eta_2^*, \ldots, \eta_6^*)}{t} \\
		&= \frac{1}{\lvert\Omega\rvert} \int_{\Omega} -2\gamma(\eta_5^*\eta_6^*\phi_4 + \eta_1\eta_2^*\phi_4) \,{\rm d}\bm{r}\\
		&\quad + \frac{1}{\lvert\Omega\rvert} \int_{\Omega} 6v(\eta_1\eta_3^*\eta_5^*\phi_4 + \eta_3\eta_2^*\eta_6^*\phi_4) \,{\rm d}\bm{r} \\
		&= \left(-2\gamma(\eta_5^*\eta_6^* + \eta_1\eta_2^*) + 6v(\eta_1\eta_3^*\eta_5^* + \eta_3\eta_2^*\eta_6^*), \phi_4\right).
	\end{align*}
	Thus, we find
	\begin{equation*}
		\frac{\delta f^{\rm bcc}}{\delta \eta_4^*} = -2\gamma(\eta_5^*\eta_6^* + \eta_1\eta_2^*) + 6v(\eta_1\eta_3^*\eta_5^* + \eta_3\eta_2^*\eta_6^*).
	\end{equation*}
	In general,
	\begin{equation}
		\frac{\delta f^{\rm bcc}}{\delta \eta_l^*} = -2\gamma(\eta_m^*\eta_n^* + \eta_i\eta_j^*) + 6v(\eta_i\eta_k^*\eta_m^* + \eta_k\eta_j^*\eta_n^*).
		\label{eqn:fs456}
	\end{equation}
	In this case, all the equations for the amplitudes in \eqref{eqn:fs123} and \eqref{eqn:fs456} are obtained by permutations of the groups $(i,j,k) = (1,2,3)$ and $(l,m,n) = (4,5,6)$.
\end{enumerate}

\subsection{The calculation of \texorpdfstring{$\phi_0$}{\phi_0}}
\label{sec:sub:phi0}

Let $\phi_j = \phi_0$ represent the amplitude of the equilibrium crystal. By assuming the amplitudes for all $\eta_j$ to be real and equal, we can set $\eta_j = \phi_0$. This can be determined by minimizing the energy functional given in \eqref{eqn:energyfunctionalineta}.

\begin{enumerate}
	\item For two-dimensional triangular lattices with $M=3$,
	\begin{align*}
		f^{s} = f^{s}_{\rm{tri}} = -2\gamma(\eta_1\eta_2\eta_3 + 
		{\rm c.c.}).
	\end{align*}
	At this point, $F$ defined by \eqref{eqn:Fexpan} becomes
	\begin{equation}
		F = \frac{1}{\lvert\Omega\rvert}\int_{\Omega} \left(\frac{45}{2}v\phi_0^4 - 4\gamma\phi_0^3 + 3B_0\phi_0^2\right)\,{\rm d}\bm{r}.
		\label{eqn:Fphi2D}
	\end{equation}
    According to the necessary condition for extrema of differentiable functions, the minimum value is obtained when
	\begin{equation}
		\phi_0 = \frac{\gamma + \sqrt{\gamma^2 - 15v B_0}}{15v}.
	\end{equation}
	
	\item For the body-centered cubic (BCC) lattice with $M=6$, 
	\begin{align*}
		f^{s} = f^{s}_{\rm bcc} = -&2\gamma(\eta_1^*\eta_2\eta_3 + \eta_2^*\eta_3\eta_5 + \eta_3^*\eta_1\eta_6 + \eta_4^*\eta_5^*\eta_6^* + {\rm c.c.})\\ 
		&+ 6v(\eta_1\eta_3^*\eta_4^*\eta_5^* + \eta_2\eta_1^*\eta_5^*\eta_6^* + \eta_3\eta_2^*\eta_6^*\eta_4^* + {\rm c.c.}).
	\end{align*}
	At this point, $F$ defined by \eqref{eqn:Fexpan} becomes
	\begin{equation}
		F = \frac{1}{\lvert\Omega\rvert}\int_{\Omega} \left(135v\phi_0^4 - 16\gamma\phi_0^3 + 6B_0\phi_0^2\right)\,{\rm d}\bm{r}.
		\label{eqn:Fphi3D}
	\end{equation}
	According to the necessary condition for extrema of differentiable functions, the minimum value is obtained when
	\begin{equation}
		\phi_0 = \frac{2\gamma + \sqrt{4\gamma^2 - 45vB_0}}{45v}.
	\end{equation}
\end{enumerate}

\subsection{Proofs of the Remarks in Section~\ref{sec:APFC Model}}
\label{apd:sub:proofs}
In this section, we provide the detailed proofs of the Remarks presented in Section~\ref{sec:APFC Model}. 

\begin{customproof}{remark1}
	\begin{align*}
		&\quad\Big\lvert \frac{1}{\varepsilon^d}\int_{C_{\varepsilon}}\eta(\bm{x})e^{\imath\bm{k}_j\cdot \frac{\bm{x}}{\varepsilon}} \,{\rm d}\bm{x} - \frac{1}{\varepsilon^d}\int_{C_{\varepsilon}}I_{\eta}e^{\imath\bm{k}_j\cdot \frac{\bm{x}}{\varepsilon}} \,{\rm d}\bm{x}\Big\rvert 
		\\
		&= \Big\lvert \int_{C_1} (\eta(\varepsilon \bm{y}) - I_{\eta}) e^{\imath\bm{k}_j\cdot \bm{y}} \,{\rm d}\bm{y}\Big\rvert \\
		&\leq \Big\lvert \Big( \int_{C_1}\lvert \eta(\varepsilon \bm{y}) - I_{\eta} \rvert^2 \,{\rm d}\bm{y} \Big)^{\frac{1}{2}} \Big( \int_{C_1}\lvert e^{\imath\bm{k}_j\cdot \bm{y}}\rvert^2 \,{\rm d}\bm{y} \Big)^{\frac{1}{2}} \Big\rvert \\
		&= \Big\lvert \Big( \int_{C_1}\big\lvert \eta(\varepsilon \bm{y}) - I_{\eta} \big\rvert^2 \,{\rm d}\bm{y} \Big)^{\frac{1}{2}}\Big\rvert \\
		&= \Big\lvert \Big( \frac{1}{\varepsilon^d}\int_{C_{\varepsilon}}\lvert \eta(\bm{x}) - \frac{1}{\varepsilon^d} \int_{C_{\varepsilon}} \eta(\bm{x})\,{\rm d}\bm{x} \rvert^2 \,{\rm d}\bm{x} \Big)^{\frac{1}{2}}\Big\rvert \\
		&\leq \Big\lvert K_{\varepsilon} \lVert \nabla \eta(\bm{x})\rVert_{L^2(C_{\varepsilon})}\Big\rvert \\
		&= O(\varepsilon).
	\end{align*}
\end{customproof}

\begin{customproof}{remark2}
	For any $\varepsilon > 0$, let $\delta' > 0$ and $\delta > 0$ such that $0 < \lvert \bm{k} \rvert < \delta'$ and $0 < \lvert -\bm{k}^2 - 2\bm{k}_j\cdot\bm{k} \rvert < \delta$. Then,
	\begin{align*}
		&\quad\Big\lvert\frac{1}{\varepsilon^d}\int_{C_{\varepsilon}} ( \nabla^2 + 2\imath\bm{k}_j\cdot\nabla )\widehat{\eta}_je^{\imath\bm{k}\cdot\frac{\bm{x}}{\varepsilon}} \,{\rm d}\bm{x}\Big\rvert\\ 
		&= \Big\lvert\int_{C_1} ( \nabla^2 + 2\imath\bm{k}_j\cdot\nabla )\widehat{\eta}_j e^{\imath\bm{k}\cdot \bm{y}} \,{\rm d}\bm{y}\Big\rvert \\
		&= \Big\lvert\int_{C_1} ( -\bm{k}^2 - 2\bm{k}_j\cdot\bm{k})\widehat{\eta}_j e^{\imath\bm{k}\cdot \bm{y}} \,{\rm d}\bm{y}\Big\rvert \\
		&\leq \Big\lvert \Big( \int_{C_1} \lvert( -\bm{k}^2 - 2\bm{k}_j\cdot\bm{k})\widehat{\eta}_j\rvert^2 \,{\rm d}\bm{y}\Big)^{\frac{1}{2}} \Big( \int_{C_1} \lvert e^{\imath\bm{k}\cdot \bm{y}}\rvert^2  \,{\rm d}\bm{y} \Big)^{\frac{1}{2}} \Big\rvert \\
		&= \Big\lvert \Big( \frac{1}{\varepsilon^d}\int_{C_{\varepsilon}} \lvert( -\bm{k}^2 - 2\bm{k}_j\cdot\bm{k})\widehat{\eta}_j\rvert^2 \,{\rm d}\bm{x} \Big)^{\frac{1}{2}}\Big\rvert \\
		&\leq \Big\lvert K_{\varepsilon} ( -\bm{k}^2 - 2\bm{k}_j\cdot\bm{k})\lVert \nabla \widehat{\eta}_j\rVert_{L^2(C_{\varepsilon})}\Big\rvert \\
		&< \Big\lvert K_{\varepsilon} \delta \lVert \nabla \widehat{\eta}_j\rVert_{L^2(C_{\varepsilon})}\Big\rvert \\
		&= O(\varepsilon).
	\end{align*}
\end{customproof}

\section{Finite Element Methods for Solving~\eqref{eqn:apfcdynamics}}
\label{sec:sub:fem}

Although we use the Fourier pseudospectral method to solve the APFC model in this work, we include a brief overview of the commonly used finite element method for completeness. This overview, based on previous studies~\cite{2017Controlling, wang2021posteriori, ortner2023framework}, focuses on spatial discretization and adaptivity techniques. For the discretization of the equation \eqref{eqn:apfcdynamics}, a conforming triangulation $\mathcal{T}_h$ of the computational domain $\Omega$ is considered, usually with simplex elements $S \in \mathcal{T}_h$ of characteristic size $h$. In the simulation of APFC, linear elements are adopted. Considering a ($\mathbb{P}_1$) FEM space, we define 
\begin{equation*}
	\mathcal{V}_h^1 = \{v \in C(\Omega, \mathbb{R}) : v\lvert_S \in \mathbb{P}_1(S, \mathbb{R}), \, S \in \mathcal{T}_h\}.
\end{equation*}
A function $y \in \mathcal{V}_h^1$ can be expressed in terms of a basis expansion: $y = \sum_{i} Y_i e_i$, 
with real coefficients $Y_i$ and basis $\{e_i\}$ of $\mathcal{V}_h^1$. A key feature of the FEM approach is splitting equation \eqref{eqn:apfcdynamics} into two second-order equations for $\partial \eta_j / \partial t$ and $\rho_j = \varsigma_j \eta_j$:
\begin{equation}
	\begin{aligned}
		\frac{\partial \eta_j}{\partial t} &= -\lvert\bm{k}_j\rvert^2 \left( B_x \varsigma_j \rho_j + B_0 \eta_j + 3v \left( A^2 - \lvert\eta_j\rvert^2 \right) \eta_j + \frac{\delta f_s}{\delta \eta_j^*} \right), \\
		\rho_j &= \varsigma_j \eta_j = \nabla^2 \eta_j + 2 \imath \bm{k}_j \cdot \nabla \eta_j.
	\end{aligned}
	\label{eqn:slip_two_eqns}
\end{equation}

Considering that the amplitude functions are complex-valued, we may use complex coefficients with real basis functions. The functions are split into real and imaginary parts and divided accordingly for numerical solution. The following array of functions is considered: 
$\alpha = \left[ \mathrm{Re}(\eta_1), \mathrm{Im}(\eta_1), \ldots, \mathrm{Re}(\eta_M), \mathrm{Im}(\eta_M) \right]$, indexed by $p = 1, \ldots, 2M$. The problem to solve then reads: for $t \in [0, T]$, find $\eta_j = \alpha_k + i\alpha_{k+1}$ and $\varsigma_j \eta_j = \zeta_k + i\zeta_{k+1}$, $k = 2j - 1$, with $\alpha_k, \alpha_{k+1}, \zeta_k, \zeta_{k+1} \in \mathcal{V}_h^1$, such that
\begin{equation}
	\begin{aligned}
		\frac{\partial \alpha_k}{\partial t} &= -\lvert\bm{k}_j\rvert^2 \bigg[ B_0 \alpha_k + B_x \left( \nabla^2 \zeta_k - 2 \bm{k}_j \cdot \nabla \zeta_{k+1} \right) \\
		&\quad + 3v \left( A^2 - \lvert\eta_j\rvert^2 \right) \alpha_k + \mathrm{Re} \left( \frac{\delta f_s}{\delta \eta_j^*} \right) \bigg], \\
		\frac{\partial \alpha_{k+1}}{\partial t} &= -\lvert\bm{k}_j\rvert^2 \bigg[ B_0 \alpha_{k+1} + B_x \left( \nabla^2 \zeta_{k+1} + 2 \bm{k}_j \cdot \nabla \zeta_k \right) \\
		&\quad + 3v \left( A^2 - \lvert\eta_j\rvert^2 \right) \alpha_{k+1} + \mathrm{Im} \left( \frac{\delta f_s}{\delta \eta_j^*} \right) \bigg], \\
		\zeta_k &= \nabla^2 \alpha_k - 2 \bm{k}_j \cdot \nabla \alpha_{k+1}, \\
		\zeta_{k+1} &= \nabla^2 \alpha_{k+1} + 2 \bm{k}_j \cdot \nabla \alpha_k.
	\end{aligned}
	\label{eqn:coffeq}
\end{equation}

The time derivatives are approximated by $\partial \alpha_k / \partial t = (\alpha_k^{n+1} - \alpha_k^n) / \Delta t_n$ and $\partial \zeta_k / \partial t = (\zeta_k^{n+1} - \zeta_k^n) / \Delta t_n$, with $\Delta t_n = t_{n+1} - t_n$ being the time step, and $n \in \mathbb{N}_0$ the index labeling time steps. The time discretization is achieved through a semi-implicit scheme, where the linear terms in \eqref{eqn:coffeq} are evaluated implicitly and the nonlinear terms explicitly.

Notice that $M$ coupled systems must be solved concurrently, with $M$ being the number of independent amplitudes according to the considered lattice symmetry. The adopted semi-implicit integration scheme in matrix form reads:
\begin{equation*}
	\mathbf{L} \cdot \mathbf{x} = \mathbf{R},
\end{equation*}
with
\begin{align}
	\mathbf{L} &= \begin{bmatrix}
		-\nabla^2 & \mathcal{A} & 1 & 0 \\
		-\mathcal{A} & -\nabla^2 & 0 & 1 \\
		G_1(\{\alpha_i^{(n)}\}) & 0 & \mathcal{K} \nabla^2 & -\mathcal{K} \mathcal{A} \\
		0 & G_2(\{\alpha_i^{(n)}\}) & \mathcal{K} \mathcal{A} & \mathcal{K} \nabla^2
	\end{bmatrix}, \\
	\mathbf{x} &= \begin{bmatrix}
		\alpha_k^{(n+1)} \\
		\alpha_{k+1}^{(n+1)} \\
		\zeta_k^{(n+1)} \\
		\zeta_{k+1}^{(n+1)}
	\end{bmatrix}, \quad 
	\mathbf{R} = \begin{bmatrix}
		0 \\
		0 \\
		H_1(\{\alpha_i^{(n)}\}) \\
		H_2(\{\alpha_i^{(n)}\})
	\end{bmatrix},
\end{align}
where $\mathcal{A} = 2 \bm{k}_j \cdot \nabla$ and $\mathcal{K} = \lvert\bm{k}_j\rvert^2 B_x$. The functions evaluated explicitly at time $t_n$ are given by
\begin{equation}
	\begin{aligned}
		G_1(\{\alpha_i\}) &= \frac{1}{\Delta t_n} + \lvert\bm{k}_j\rvert^2 B_0 + 3v \lvert\bm{k}_j\rvert^2 \left( A^2 + \alpha_k^2 - \alpha_{k+1}^2 \right), \\
		G_2(\{\alpha_i\}) &= \frac{1}{\Delta t_n} + \lvert\bm{k}_j\rvert^2 B_0 + 3v \lvert\bm{k}_j\rvert^2 \left( A^2 + \alpha_{k+1}^2 - \alpha_k^2 \right), \\
		H_1(\{\alpha_i\}) &= \left[ \frac{1}{\Delta t_n} + 6 \lvert\bm{k}_j\rvert^2 v \alpha_k^2 \right] \alpha_k - \lvert\bm{k}_j\rvert^2 \mathrm{Re} \left( \frac{\delta f_s}{\delta \eta_j^*} \right), \\
		H_2(\{\alpha_i\}) &= \left[ \frac{1}{\Delta t_n} + 6 \lvert\bm{k}_j\rvert^2 v \alpha_{k+1}^2 \right] \alpha_{k+1} - \lvert\bm{k}_j\rvert^2 \mathrm{Im} \left( \frac{\delta f_s}{\delta \eta_j^*} \right).
	\end{aligned}
	\label{eqn:linearization}
\end{equation}

Furthermore, spatial adaptivity can be adopted to improve the efficiency of numerical simulations. There are many methods to choose from, such as those based on error estimates or indicators. Due to the presence of defects, the amplitude functions may oscillate differently. Based on this property, numerical methods have been developed that focus on the oscillation of the complex amplitudes. We find the region where the quantity $\lVert \nabla A^2 \rVert_F$ exceeds an arbitrary threshold, where $\lVert \cdot \rVert_F$ denotes the Frobenius norm. The method then considers a refinement of the computational grid to ensure proper characterization of all the $\eta_j$ functions.

\begin{acknowledgements}
We are deeply grateful to Prof.~Siran Li (Shanghai Jiao Tong University) for insightful discussions and valuable suggestions that greatly improved this paper, particularly for drawing our attention to the analysis of the analyticity of the APFC model and for pointing us to several helpful references on this topic.
\end{acknowledgements}

\section*{Funding}
X.~Wei, Y.~Wang and L.~Zhang are partially supported by the National Natural Science Foundation of China (No. 12271360). K.~Jiang is partially supported by the National Key Research and Development Program of China (No. 2023YFA1008802), the National Natural Science Foundation of China (No. 12171412), the Science and Technology Innovation Program of Hunan Province (No. 2024RC1052) and the Innovative Research Group Project of Natural Science Foundation of Hunan Province of China (No. 2024JJ1008). Y.~Wang is also partially supported by the Development Postdoctoral Scholarship for Outstanding Doctoral Graduates from Shanghai Jiao Tong University. L.~Zhang is also partially supported by the Shanghai Municipal Science and Technology Project (No. 23JC1402300) and the Fundamental Research Funds for the Central Universities.

\section*{Data Availability}
Data sharing is not applicable to this paper as no datasets were generated or analyzed during the current study.

\section*{Conflict of interest}
The authors declare that they have no conflict of interest.

\bibliographystyle{spmpsci}      
\bibliography{ref}
\end{document}